\documentclass[reprint,superscriptaddress,amssymb,amsmath,sn,nature]{revtex4-2}

\usepackage{graphicx}
\usepackage{amssymb}
\usepackage{epstopdf}
\usepackage{upgreek}
\usepackage{dcolumn}
\usepackage{bm}
\usepackage{gensymb}
\usepackage{braket}
\usepackage{amsmath}
\usepackage{mathtools}
\usepackage{tikz}
\usepackage[colorlinks=true,allcolors=blue]{hyperref}

\usepackage{array}
\newcolumntype{C}[1]{>{\centering\arraybackslash}m{#1}}

\usepackage{xcolor}

\usepackage[resetlabels,labeled]{multibib}

\usepackage{xfrac}

\usepackage{comment}

\usepackage{hhline}

\usepackage[normalem]{ulem}

\allowdisplaybreaks

\DeclareMathOperator{\arccosh}{arccosh}

\makeatletter
\providecommand{\killfloatstyle}{}
\providecommand{\FR@redefs}{}
\providecommand{\flrow@setlist}[1]{}
\providecommand{\FRifFBOX}{}
\providecommand{\@@setframe}[1]{}
\providecommand{\@@FStrue}{}
\makeatother

\expandafter\ifx\csname package@font\endcsname\relax\else
\expandafter\expandafter
\expandafter\usepackage
\expandafter\expandafter
\expandafter{\csname package@font\endcsname}%
\fi
\newcommand{\ie}{\emph{i.e.}}

\newcommand{\microamp}{$\upmu$A}

\newcommand{\micron}{$\upmu$m~}

\newcommand{\be}{\begin{eqnarray}}
\newcommand{\ee}{\end{eqnarray}}
\newcommand{\bfig}{\begin{figure}}
	\newcommand{\efig}{\end{figure}}

\definecolor{myorange}{rgb}{0.97,0.59,0.27}

\DeclareFontFamily{U}{mathb}{}
\DeclareFontShape{U}{mathb}{m}{n}{
	<-5.5> mathb5
	<5.5-6.5> mathb6
	<6.5-7.5> mathb7
	<7.5-8.5> mathb8
	<8.5-9.5> mathb9
	<9.5-11.5> mathb10
	<11.5-> mathbb12
}{}

\begin{document}
\makeatletter
  \@namedef{figure}{\killfloatstyle\def\@captype{figure}\FR@redefs
    \flrow@setlist{{figure}}%
    \columnwidth\columnwidth\edef\FBB@wd{\the\columnwidth}%
    \FRifFBOX\@@setframe\relax\@@FStrue\@float{figure}}%
\makeatother

\title{A Traveling-Wave Parametric Amplifier and Converter}

\author{M. Malnou}
\email{maxime.malnou@nist.gov}
\affiliation{National Institute of Standards and Technology, 325 Broadway, Boulder, CO 80305, USA}
\affiliation{University of Colorado, 2000 Colorado Ave., Boulder, CO 80309, USA}
\author{B. T. Miller}
\affiliation{National Institute of Standards and Technology, 325 Broadway, Boulder, CO 80305, USA}
\affiliation{University of Colorado, 2000 Colorado Ave., Boulder, CO 80309, USA}
\author{J. A. Estrada}
\affiliation{National Institute of Standards and Technology, 325 Broadway, Boulder, CO 80305, USA}
\affiliation{University of Colorado, 2000 Colorado Ave., Boulder, CO 80309, USA}
\author{K. Genter}
\affiliation{National Institute of Standards and Technology, 325 Broadway, Boulder, CO 80305, USA}
\affiliation{University of Colorado, 2000 Colorado Ave., Boulder, CO 80309, USA}
\author{K. Cicak}
\affiliation{National Institute of Standards and Technology, 325 Broadway, Boulder, CO 80305, USA}
\author{J. D. Teufel}
\affiliation{National Institute of Standards and Technology, 325 Broadway, Boulder, CO 80305, USA}
\affiliation{University of Colorado, 2000 Colorado Ave., Boulder, CO 80309, USA}
\author{J. Aumentado}
\affiliation{National Institute of Standards and Technology, 325 Broadway, Boulder, CO 80305, USA}
\author{F. Lecocq}
\email{florent.lecocq@nist.gov}
\affiliation{National Institute of Standards and Technology, 325 Broadway, Boulder, CO 80305, USA}
\affiliation{University of Colorado, 2000 Colorado Ave., Boulder, CO 80309, USA}
\date{\today}

\begin{abstract} 

High-fidelity qubit measurement is a critical element of all quantum computing architectures. In superconducting systems, qubits are typically measured by probing a readout resonator with a weak microwave tone that must be amplified before reaching the room temperature electronics. Superconducting parametric amplifiers have been widely adopted as the first amplifier in the chain, primarily because of their low noise performance, approaching the quantum limit. However, they require isolators and circulators to route signals up the measurement chain and to protect qubits from amplified noise. While these commercial components are wideband and simple to use, their intrinsic loss, size, and magnetic shielding requirements impact overall measurement efficiency and scalability. Here we report a parametric amplifier that achieves both broadband forward amplification and backward isolation in a single, compact, non-magnetic circuit that could be integrated on chip with superconducting qubits. The approach relies on a nonlinear transmission line that supports traveling-wave parametric amplification of forward propagating signals, and isolation via frequency conversion of backward propagating signals. This traveling-wave parametric amplifier and converter has the potential to reduce the readout hardware overhead when scaling up the size of superconducting quantum computers.

\end{abstract}

\maketitle 

A standard method for assessing the state of a superconducting qubit is to dispersively couple it to a resonator and measure a state-dependent frequency shift in the resonator's microwave response \cite{Blais2021_cQED}. This dispersive interaction allows for an approximate quantum nondemolition (QND) qubit readout, so long as the readout power is low enough, on the order of a few photons. This read-out drive must then be amplified over six orders of magnitude to overwhelm the noise of room-temperature electronics, with minimal degradation of its signal-to-noise ratio. At the same time, in the dispersive measurement scheme the qubit frequency becomes dependent on the number of photons in the readout resonator, leading to a qubit dephasing rate proportional to the cavity occupancy \cite{Wang2019cavity}. This imposes stringent simultaneous requirements on the amplification chain to have large gain and low added noise in the forward direction, while maintaining no excess noise in the backward direction.

Typical amplification chains for qubit readout use parametric amplifiers as the first stage of gain, due to noise performance fundamentally limited by Heisenberg’s uncertainty principle \cite{Caves1982quantum,Aumentado2020superconducting}. However, to enforce directionality and protect the readout cavity against amplified noise, several microwave circulators or isolators must be placed between the readout resonator and the amplifier \cite{Arute2019quantum}. In addition to their physical size and high magnetic fields being obstacles for scaling, they also introduce loss, which degrades the system noise performance.

To circumvent these issues, parametric alternatives to the Faraday effect used in conventional circulators have been investigated \cite{Kamal2011Noiseless,Metelmann2015nonreciprocal,Ranzani2019circulators}. Recent developments include building parametric circulators and directional amplifiers \cite{Sliwa2015Reconfigurable,lecocq2017nonreciprocal,Chapman2017widely}, and applying them to perform efficient qubit measurements \cite{Abdo2014Josephson,Lecocq2021efficient,Rosenthal2021efficient}. While promising, these various devices remain relatively narrow-band and difficult to operate, hindering their large scale deployment in state-of-the-art quantum computers. Another alternative is to use traveling-wave parametric amplifiers (TWPAs) \cite{Esposito2021perpective}, in which signals are amplified while copropagating with a strong pump in a nonlinear transmission line. As such, they are inherently directional, in addition to having high bandwidth, high saturation power, and near quantum-limited noise performance \cite{macklin2015near,planat2020photonic,malnou2021three}. However, in practice, small impedance mismatches are unavoidable, leading to pump and noise reflections that turn into backward gain, or more generally excess noise traveling back to the device under test. Consequently, isolators, along with their drawbacks, are still required between the readout cavities and traditional TWPAs.

\begin{figure}[h!]
    \vspace{1pt}
	\includegraphics[scale=0.95]{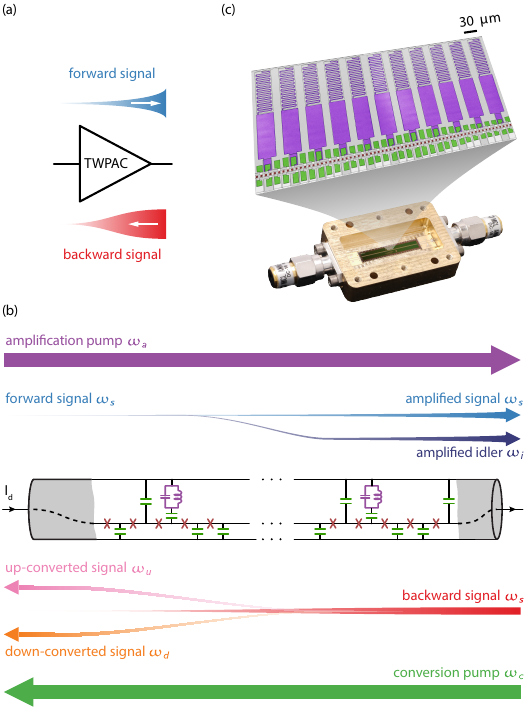}
    \caption{The concept of a traveling-wave parametric amplifier and converter (TWPAC). (a) Signals propagating from left to right (forward direction) are amplified, while signals propagating from right to left (backward direction) are converted out of the band. (b) The TWPAC consists of a current-biased nonlinear lumped-element transmission line. Each unit cell consists of a series inductor made with a Josephson junction (brown) shorted to ground by a linear capacitor (green), whose value is periodically modulated (see main text). Resonators to ground (purple) are added every six cells which, together with the periodically varying capacitors, enable phase-matching for the two three-wave mixing parametric processes: (i) from left to right, forward parametric amplification, whereby a pump (purple arrow) transfers its energy to a co-propagating signal (blue) and to its corresponding idler (dark blue). (ii) From right to left, frequency conversion, whereby a pump (green arrow) mixes a signal (red)  up (pink) or down (orange), away from the signal frequency band. (c) Experimental implementation of a TWPAC in a $2$\,cm-long device, comprising 2640 unit cells. The close-up micrograph shows, in false color, the junctions (brown), ground capacitors (green) and tank circuits (purple).}
    \label{fig:concept}
\end{figure}

In this Article, we report the development of a TWPA that incorporates forward gain with backward isolation (see Fig.\,\ref{fig:concept}a). The latter is achieved via frequency conversion of backward propagating signals \cite{Ranzani2017wideband} instead of the Faraday effect, making our traveling-wave parametric amplifier and converter (TWPAC) directly integrable with superconducting qubits. We experimentally demonstrate the simultaneous forward amplification and backward isolation of signals over a $500$\,MHz usable bandwidth, with a gain of about $7$\,dB, and isolation as low as $-20$\,dB. We characterize the noise performance of the system with a shot-noise tunnel junction (SNTJ). Using the TWPAC as the first amplifier, we achieve a system-added noise of $5.2$ quanta in the operation band, on par with the measured performance of commonly used TWPA-based amplifier chains \cite{macklin2015near,malnou2024low}. 

\subsection*{Concept and design}

The TWPAC consists of a nonlinear transmission line that is current-biased to support two counter-propagating three-wave mixing parametric processes, see Fig.\,\ref{fig:concept}b. In the forward direction, a strong pump at frequency $\omega_a$ is used to amplify a co-propagating signal at frequency $\omega_s$ and an idler at frequency $\omega_i = \omega_a - \omega_s$. In the backward direction, using a second strong pump at frequency $\omega_c$, a signal at frequency $\omega_s$ is either down-converted or up-converted to frequencies $\omega_d = \omega_s-\omega_c$ or $\omega_u = \omega_s + \omega_c$, respectively. This frequency conversion effectively swaps any states contained within the signal band with vacuum states coming from the up- or down-converted frequency bands, thus providing the isolation property of the TWPAC. Importantly, both the parametric amplification (PA) and frequency conversion (FC) processes are directional, because each of these processes must respect a specific phase-matching condition that is predominantly achieved for signals co-propagating with the pumps (see supplementary information, section Ic).

We realize this nonlinear transmission line using Josephson junctions, see Fig.\,\ref{fig:concept}b. Each unit cell within this line is composed of a series inductor from a single niobium trilayer Josephson junction ($I_c\approx 5$\,\microamp) and a capacitor to ground from a parallel-plate using a low-loss amorphous silicon dielectric ($\tan{\delta}\approx4\times10^{-4}$) \cite{lecocq2017nonreciprocal}. The total device comprises $2650$ unit cells to achieve a $\approx2$\,cm long transmission line, see Fig.\,\ref{fig:concept}c. In the signal band, the line has a $\approx50$\,\ohm{} characteristic impedance and a phase velocity $\approx2\%$ of the speed of light in vacuum. The TWPAC detailed layout is available in Fig.\,\ref{fig:TWPAClayout}.

\begin{figure}[h!]
    \vspace{1pt}
	\includegraphics[scale=0.95]{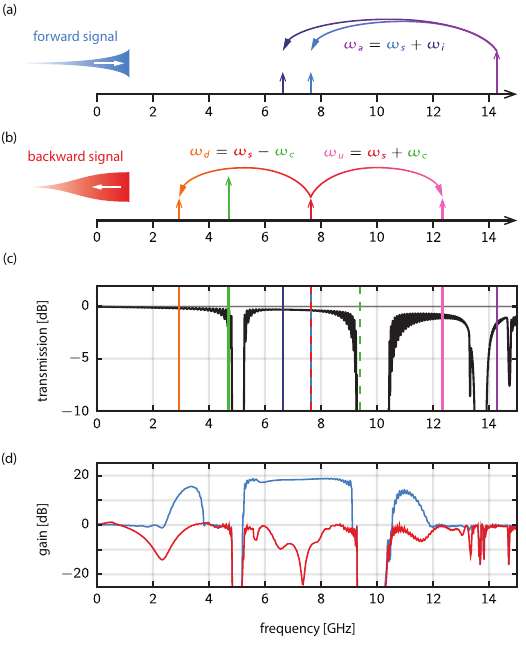}
    \caption{The theoretical model supporting the TWPAC's design. (a) In the forward (\ie{} left to right) direction, pump photons at a frequency $\omega_a/2\pi\approx14$\,GHz are used to amplify a signal at frequency $\omega_s/2\pi\approx7.5$\,GHz, while also amplifying an idler at frequency $\omega_i=\omega_a-\omega_s$. (b) In the backward (\ie{} right to left) direction, the signal gets up-converted at $\omega_u=\omega_s+\omega_c$, or down-converted at $\omega_d=\omega_s-\omega_c$ using pump photons at $\omega_c/2\pi\approx4.75$\,GHz. (c) The TWPAC's calculated transmission as a function of frequency is flat and close to $0$\,dB, except at distinct stopbands. The highest-frequency stopband allows us to match the phase of the PA pump (purple line) with the summed phases of the signal (blue) and idler (dark blue) that lie in the $6$ to $8$\,GHz band. The lowest frequency stopband allows us to achieve phase matching between the conversion pump (green), the signal (dashed red) and the down-converted signal (orange). The up-converted signal (pink) can also be created via frequency conversion, albeit not phase-matched. Ideally, the second harmonic of the conversion pump (dashed green) falls into the second stopband. (d) We solve the equations coupling all these modes (see the supplementary information, section Ib) to calculate their steady-state amplitude as a function of input signal frequency. Here the TWPAC is biased with a dc-current $I_d=1.5$\,\microamp{} (the junctions' critical current is $I_c=5$\,\microamp), the amplification pump frequency and power are $\omega_a/2\pi=14.27$\,GHz and $P_a=-73$\,dBm, respectively, and the conversion pump frequency and power are $\omega_c/2\pi=4.7$\,GHz and $P_c=-73$\,dBm, respectively.}
    \label{fig:design}
\end{figure}

To fulfill the PA and FC phase-matching conditions, we engineer the dispersion relation of the transmission line: (i) every six cells, a tank circuit is interposed between the capacitor and the ground, and (ii) the capacitors vary periodically along the line. These features create stopbands in the line's transmission where signals cannot propagate \cite{HoEom2012a,macklin2015near,malnou2021three}. Close to these stopbands the phase velocity varies strongly with frequency, which we exploit to obtain phase matching. We calculate the linear response of the line, in the presence of dielectric loss, to obtain the transmission as a function of frequency presented in Fig.\,\ref{fig:design}c. The tank circuits create a stopband just below $14$\,GHz, above which we place our PA pump such that signals centered around the half pump frequency ($\approx7$\,GHz) get amplified (see Fig.\,\ref{fig:design}a). The periodic variation of the capacitors to ground creates two stopbands, one just above $5$\,GHz, and one around $10$\,GHz. The FC pump is placed below the first stopband. As a result, signals around $7$\,GHz can be down-converted to $\approx3$\,GHz or up-converted to $\approx12$\,GHz (see Fig.\,\ref{fig:design}b). Ideally, the second stopband suppresses the second-harmonic generation of the conversion pump at frequency $\omega_{c2} = 2\omega_c$. A detailed discussion of the TWPAC design is available in the supplementary information, section II.

We calculate the theoretical response of the device in the presence of both pumps by solving the equations of motion for the minimal mode basis $\{a,s,i,c,d,u,c_2\}$ with frequencies $\{\omega_a,\omega_s,\omega_i,\omega_c,\omega_d,\omega_u,2\omega_c\}$. We assume here that these modes are only coupled via three-wave mixing interactions, see the supplementary information, section Ib. In these coupled-mode equations (CME), injecting a signal, either in the forward or in the backward direction allows us to calculate the TWPAC's forward and backward transmission, respectively. For a given combination of pump powers, pump frequencies and dc current bias, we achieve simultaneous forward gain and backward isolation, within a band centered around $7$\,GHz, see Fig.\,\ref{fig:design}d. The gain profile spans from about $3$\,GHz to $12$\,GHz. When the signal falls into one of the stopbands, it cannot propagate and the gain is therefore null at this frequency. Alternatively, when the idler falls into a stopband, it cannot propagate either, which results in unity gain for the associated signal. From $6$\,GHz to $8$\,GHz, this simple model predicts close to $20$\,dB of gain, concurrently with about $10$\,dB of isolation, dipping as low as $25$\,dB.

\subsection*{Measurement and analysis}

\begin{figure*}[htpb!]
    \vspace{1pt}
	\includegraphics[scale=1.0]{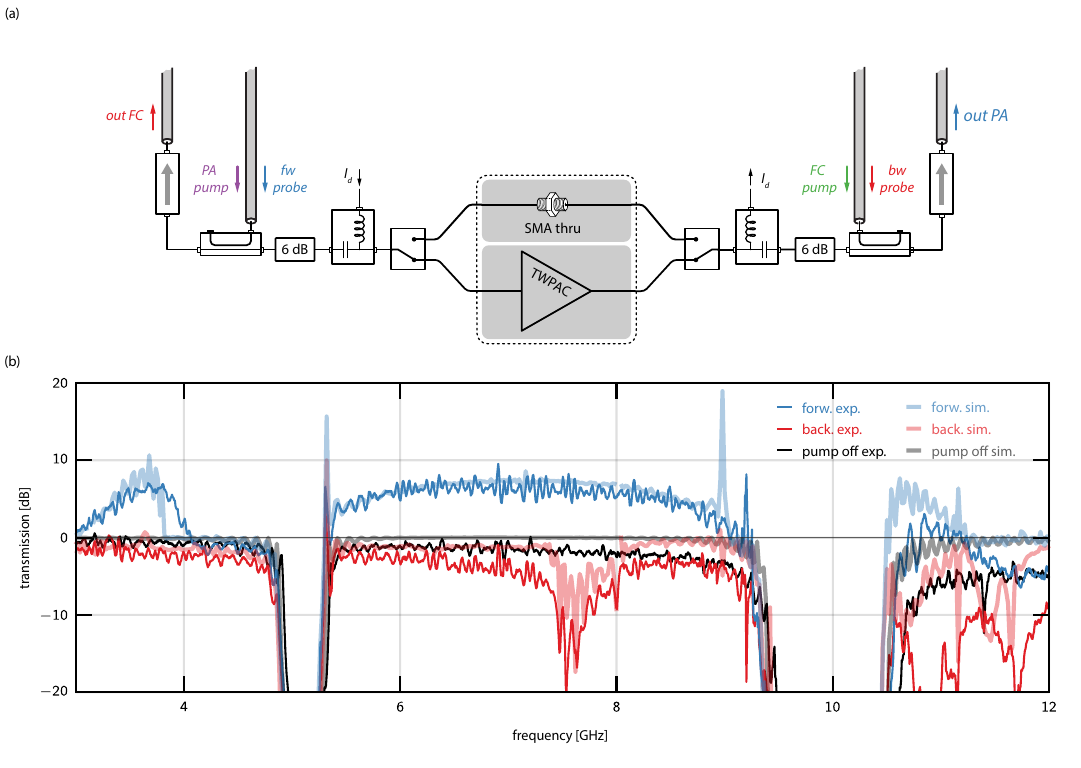}
    \caption{Experimental characterization of the TWPAC's scattering parameters. (a) We use two symmetric readout chains, between which the TWPAC is inserted. Starting from the TWPAC and extending outward in both directions: a pair of microwave switches allows us to compare the TWPAC's transmission to that of a SubMiniature version A (SMA) through connector, a pair of bias tees are used to deliver the dc current bias $I_d$, a pair of attenuators ensure that the environment is broadly matched to 50\,\ohm, a pair of $20$\,dB directional couplers deliver the pump and probe tones, a pair of isolators protect from noise and tones reflected off the HEMTs, and finally a pair of HEMTs at 4\,K (not shown) amplify the transmitted (right) and reflected (left) signals. (b) The TWPAC transmission as a function of frequency is plotted in the forward (blue) and backward (red) directions, when $I_d=1$\,\microamp, $\omega_a/2\pi=14.52$\,GHz, $P_a=-75$\,dBm, $\omega_c/2\pi=3.15$\,GHz and $P_c=-80$\,dBm (at the chip inputs), and overlapping the transmission obtained from the time-domain simulations in the forward (faded blue) and backward (faded red) directions, for $I_d=1$\,\microamp, $\omega_a/2\pi=14.3$\,GHz, $P_a=-70$\,dBm, $\omega_c/2\pi=3.14$\,GHz and $P_c=-74$\,dBm. We also plot the experimental (black) and time-domain simulated (faded black) transmission when both pumps are turned off.}
    \label{fig:scatdata}
\end{figure*}

We then test the TWPAC presented in Fig.\,\ref{fig:concept}c at cryogenic temperatures, using the experimental setup presented in Fig.\,\ref{fig:scatdata}a. Notably, this setup is symmetric in the forward and backward directions, allowing us to measure the TWPAC's complete set of scattering parameters. In addition, a pair of switches allows us to compare \textit{in situ} the TWPAC's transmission to that of a through cable. Remarkably, when current biased, the transmission, shown in black in Fig.\,\ref{fig:scatdata}b, decreases at most by $\approx2$\,dB when reaching the higher end of the operation bandwidth ($\approx8$\,GHz), thanks to the low-loss a-Si dielectric. The forward and backward transmission through the device (referenced to the cable), with the dc current bias, PA pump, and FC pump turned on, are shown in blue and red respectively in Fig.\,\ref{fig:scatdata}b. In the forward direction, the TWPAC generates about $7$\,dB of gain, spanning over close to an octave of bandwidth. Simultaneously, in the backward direction the TWPAC generates at least $5$\,dB of isolation over about $800$\,MHz of bandwidth, with sweet-spot frequencies where the isolation reaches $20$\,dB. The TWPAC input and output reflection coefficients are similar to that of the through cable, and the input $1$\,dB compression point is measured to be $\approx-90$\,dBm (see Fig.\,\ref{fig:1dbcp}).

Both the gain and isolation profiles in Fig.\,\ref{fig:scatdata}b qualitatively agree with the basic model presented in Fig.\,\ref{fig:design}d, but not quantitatively. In fact, we found the isolation to be optimal when using a FC pump frequency at $3.15$\,GHz, in contradiction with the $4.7$\,GHz, predicted by the minimal model described by the CME. To go beyond the model that guided the TWPAC design, we simulate the line's nonlinear response using a time-domain solver (WRspice \cite{WRspice,gaydamachenko2022numerical,Kissling2023vulnerability}). In contrast to the CME, this method grasps real-life non-idealities: it uses the full Josephson current-phase relation, and accounts for all propagating modes and mixing processes. The simulated circuit response, shown in faded colors in Fig.\,\ref{fig:scatdata}b, is in quantitative agreement with the experimental results, using values for the dc current bias, pump frequencies and pump powers close to their estimated experimental values. Note however that the transmission loss, for example coming from the dielectric, cannot be straightforwardly represented in a time domain simulation\cite{engin2004time}, and is therefore not included here, which explains why the simulated unpumped transmission does not fully agree with the measurement.

In light of the time-domain simulations, we revisit the CME to include all the four-wave mixing (4WM) terms that couple the modes $a,s,i,c,d,u,c_2$ with each other. These updated CME then capture the observed isolation behavior when using a FC pump frequency at $3.15$\,GHz (see the supplementary information, Fig.\,S4), which in fact corresponds to the frequency for which the up-conversion process is phase-matched. Therefore, despite not being phase-matched, the 4WM terms significantly reduce the efficacy of the down-conversion process in the TWPAC. 

While the root causes for the TWPAC's limited gain and isolation are still under investigation, potential sources include: (i) fabrication imperfection leading to parameters spread \cite{Kissling2023vulnerability}; (ii) spurious 4WM processes or 3WM amplification of $c_2$; (iii) pump depletion producing the high-gain peak (along with its idler) observed at the right edge of the first stopband. This peak may come from a runaway amplification effect due to a large impedance mismatch between the TWPAC and its environment \cite{planat2020photonic,kern2023reflection}. It is a common issue for TWPAs where the phase matching is achieved using periodic loading or other resonant techniques \cite{HoEom2012a,macklin2015near,Chaudhuri2015simulation,Chaudhuri2017broadband}. More details can be found in the methods.

\begin{figure}[h]
    \vspace{1pt}
	\includegraphics[scale=0.95]{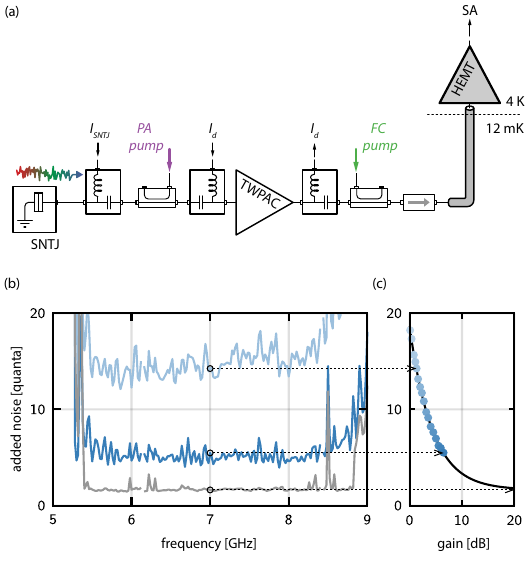}
    \caption{Characterization of the TWPAC's noise performance. (a) The amplification chain comprises the TWPAC as the first amplifier, and a calibrated noise source (SNTJ) at the chain's input. (b) The system-added noise $N_\mathrm{sys}=N_1+N_2/G$ as a function of frequency diminishes as the TWPAC's gain $G$ increases. At very low gain (light blue curve) $N_\mathrm{sys}$ is dominated by $N_2$, the effective HEMT-added noise, which includes the loss of the microwave components placed between the SNTJ and the HEMT. At the highest operable TWPAC gain (blue curve), $N_\mathrm{sys}$ drops to $5.2$ quanta on average between $5.5$ and $8.5$\,GHz. In that band, the contribution of the TWPAC and of its preceding microwave components is $N_1=1.7$ quanta (gray curve). Note that for all three curves, the noise at the frequency of the second harmonic of the FC pump ($6.3$\,GHz) and its corresponding idler ($8.22$\,GHz) cannot be properly measured, and is therefore omitted. (c) We measure $N_\mathrm{sys}$ as a function of the TWPAC gain, plotted here for the frequency bin centered around $7$\,GHz (blue dots). The fitted response (black curve) allows us to obtain $N_1$, the asymptotic system-added noise at high TWPAC gain.}
    \label{fig:noise}
\end{figure}

\subsection*{Noise performance}

Irrespective of its isolation properties, a pre-amplifier suitable for qubit readout must operate close to the quantum limit for added noise. In a separate cooldown, we measure the noise performance of a chain where the TWPAC is the first amplifier, using a shot-noise tunnel junction \cite{Spietz2003primary,spietz2006shot,malnou2024low} (SNTJ) as a calibrated noise source at the chain's input, see Fig.\,\ref{fig:noise}a. We operate the TWPAC with both PA and FC pumps turned on, whose powers and frequencies are similar to those used to capture its scattering response. Despite modest TWPAC gain, it is already sufficient to overwhelm most of the noise $N_2$ generated by the following elements in the chain, such that the average system-added noise (\ie, not including the signal input vacuum noise) is $N_\mathrm{sys}\approx5.2$ quanta between 5.5 and 8.5 GHz, see Fig.\,\ref{fig:noise}b. Varying the PA pump we measured $N_\mathrm{sys}$ at various TWPAC gains $G$, see Fig.\,\ref{fig:noise}c. We then fit the data using $N_\mathrm{sys}=N_1+N_2/G$ to infer $N_1$, the noise added by the \textit{effective} first amplifier, comprising the TWPAC, its packaging, and all the external cabling and components between itself and the SNTJ \cite{malnou2024low}. On average, between $5.5$ and $8.5$\,GHz, $N_1$ approaches $1.7$ quanta extrapolating to the high gain limit, relatively close to the quantum limit of $0.5$ quanta and consistent with the presence of loss before the TWPA chip. Note also that the dependence of $N_\mathrm{sys}$ on $G$ does not change when turning off the FC pump (see Fig.\,\ref{fig:comparison_noise}), indicating that the FC process does not alter the TWPAC's noise performance.

\subsection*{Conclusion}

We have reported the development of a traveling-wave parametric amplifier that generates backward isolation in addition to forward amplification. The TWPAC’s bandwidth, noise, and power handling are on par with traditional TWPAs, and it remains easy to operate, because it only requires two pumps. We achieved quantitative agreement between the experimental data and time-domain simulation, and qualitative agreement of both to the refined theory.

Potential  future improvements include: integrating on chip external microwave components, such as bias tees and directional couplers \cite{Howe2025compact}, or alternatively, input/output multiplexers \cite{Lecocq2020microwave}, leading to smaller device footprint, reduced insertion loss, and better noise performance; using Kerr-free nonlinear elements to mitigate spurious 4WM processes \cite{gaydamachenko2025an,Ranadive2022Kerr}; placing the conversion pump higher in frequency than the signal band to push its harmonic away from the signal and up-converted bands; and optimizing the dispersion engineering to mitigate spurious pump harmonics, unwanted parametric processes, and impedance mismatches.

With such improvements, we expect to achieve $20$\,dB of gain and isolation over more than a gigahertz of bandwidth, enabling the simultaneous readout of tens of qubits without the use of an intermediate circulator. Devices based on this concept could thus improve qubit measurement efficiency, and reduce the number of conventional microwave circulators and isolators required for the operation of large-scale quantum computers.

\section*{Methods}

\subsection{The TWPAC layout}
\label{sec:TWPAClayout}

\begin{figure*}[h]
    \vspace{1pt}
	\includegraphics[scale=0.95]{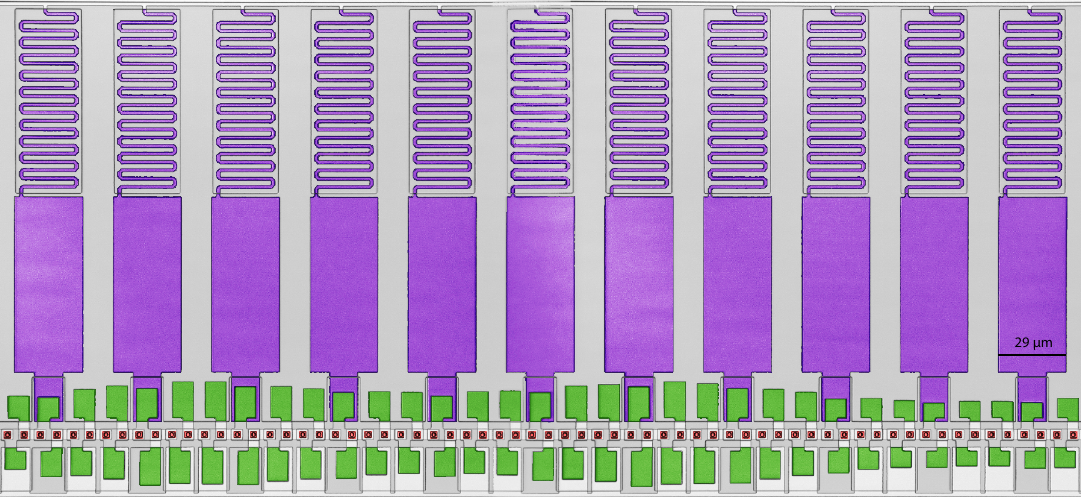}
    \caption{False color scanning-electron microscope picture of one TWPAC supercell. The color coding is identical to that of Fig.\,\ref{fig:concept}: the Josephson junctions are colored in brown, the capacitors to ground are colored in green, and the resonators to ground are colored in purple.}
    \label{fig:TWPAClayout}
\end{figure*}

Figure \ref{fig:TWPAClayout} shows one  TWPAC supercell. It is composed of $66$ cells, each containing one Josephson junction and one capacitor to ground. The resonators to ground are placed every six cells.

\subsection*{The TWPAC response to a single pump tone}

\begin{figure*}[h]
    \vspace{1pt}
	\includegraphics[scale=0.95]{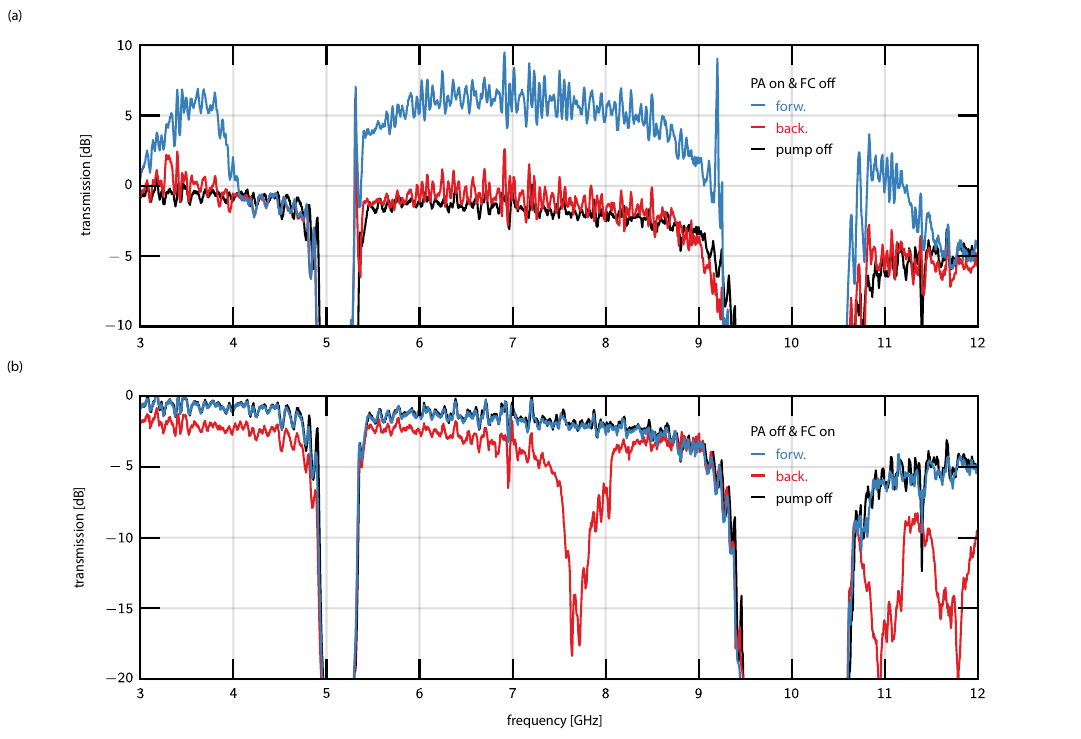}
    \caption{The gain and isolation of the TWPAC, when either the PA pump (a) or the FC pump (b) is turned on. The TWPAC biasing parameters are similar to that of Fig.\,\ref{fig:scatdata}b: $I_d=1$\,\microamp, $\omega_a/2\pi=14.52$\,GHz, $P_a=-75$\,dBm, $\omega_c/2\pi=3.15$\,GHz and $P_c=-80$\,dBm (at the chip inputs).}
    \label{fig:PAorFC}
\end{figure*}

Figure \ref{fig:PAorFC} shows the forward and backward transmission through the TWPAC, when only the PA pump is turned on (Fig.\,\ref{fig:PAorFC}a) or when only the FC pump is turned on (Fig.\,\ref{fig:PAorFC}b). It demonstrates that both parametric processes can be activated independently of each other. The pumps' powers and frequencies are similar to that used in Fig.\,\ref{fig:scatdata}b. When only the PA pump is turned on, the backward transmission is slightly higher than when the PA pump is off, in particular in the $6$ to $8$\,GHz band. It shows that in this conventional operating situation where the TWPAC functions as a conventional TWPA, there is some non-negligible reverse gain.

\subsection{Signals reflected off the TWPAC}

\begin{figure*}[h]
    \vspace{1pt}
	\includegraphics[scale=0.95]{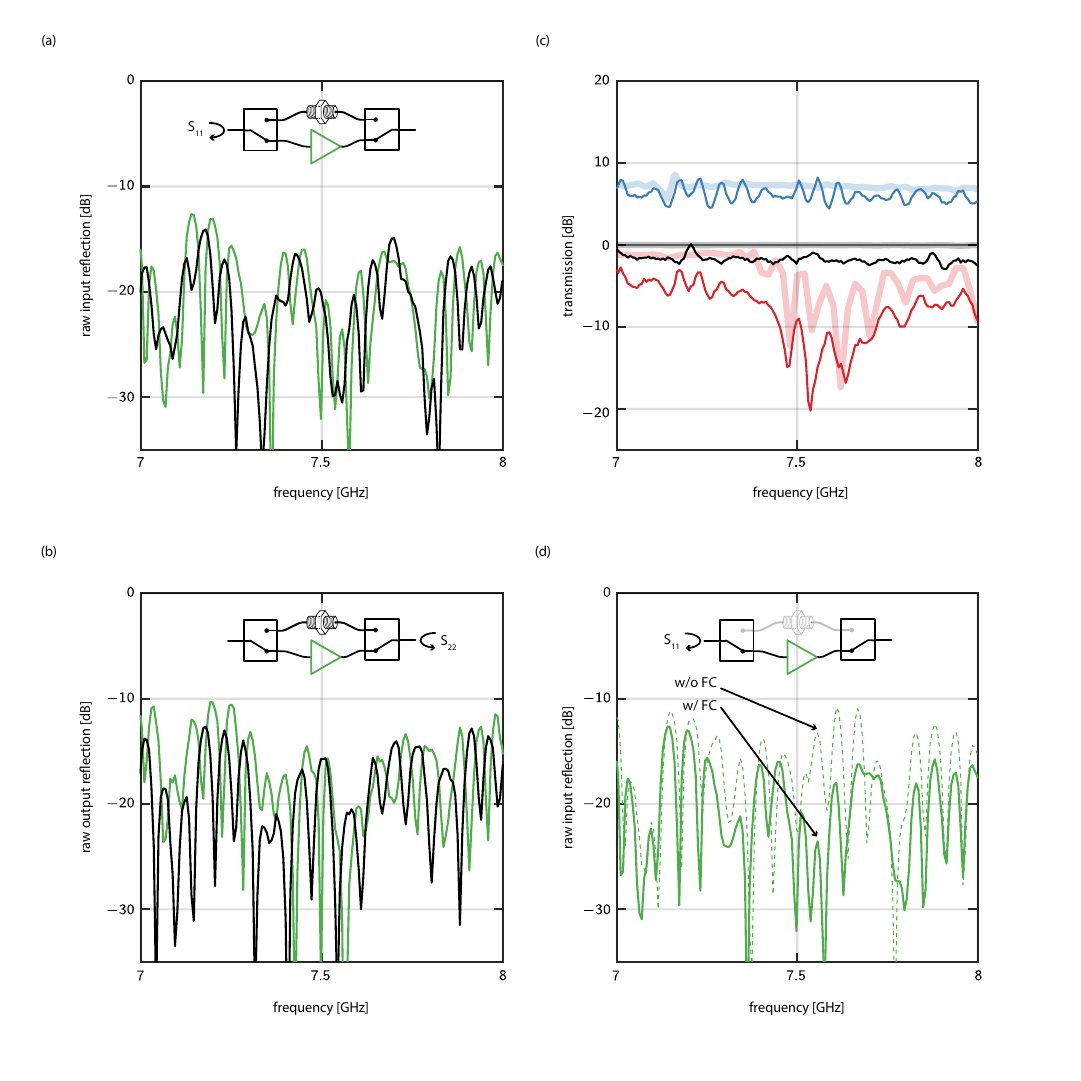}
    \caption{The full two-port scattering parameters of the TWPAC, in the frequency window where we observe both PA and FC. The TWPAC biasing parameters are similar to that of Fig.\,\ref{fig:scatdata}b. The input (a) and output (b) reflections of the through cable (black) and of the TWPAC (green) with the FC conversion pump turned on are plotted as a function of frequency. (c) The transmission as a function of frequency is the same as presented in Fig.\,\ref{fig:scatdata}b. (d) The input reflection without (dashed line) and with (plain line) the FC pump turned on. }
    \label{fig:TWPACreflection}
\end{figure*}

In the main text we presented the forward and backward transmission of the TWPAC. More precisely, we presented $10\log\lvert S_\mathrm{21} \vert^2$ and $10\log\lvert S_\mathrm{12} \vert^2$, respectively the forward and backward power transmission in decibels. Figures \ref{fig:TWPACreflection}a and b show the raw input reflections (i.e. measured by the VNA, uncorrected from the chain's gain and attenuation) $10\log\lvert S_\mathrm{11} \vert^2$ and output reflections $10\log\lvert S_\mathrm{22} \vert^2$, around the frequency window where the TWPAC presents forward amplification and backward isolation (Fig.\,\ref{fig:TWPACreflection}c), when both the PA and FC pump are turned on. They are as good as that of the through cable. Also, in that same band the input reflection with the FC pump turned off is degraded, compared to when it is on (Fig.\,\ref{fig:TWPACreflection}d).

\subsection{The TWPAC saturation}

\subsubsection{The $1$\,dB compression point}
\label{sec:P1db}

\begin{figure*}[h]
    \vspace{1pt}
	\includegraphics[scale=0.95]{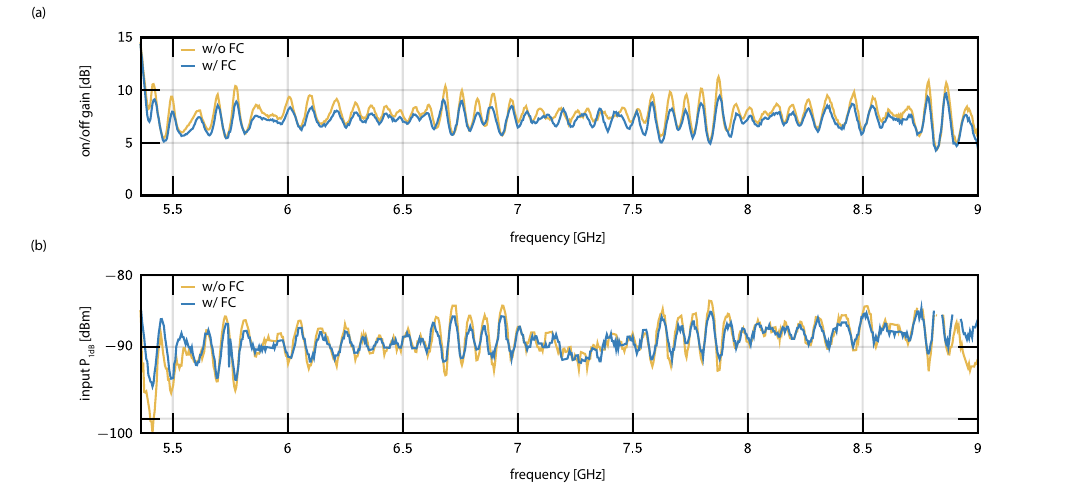}
    \caption{The TWPAC saturation power, focusing on the frequency window where we observed both PA and FC. (a) The TWPAC gain as small VNA probe tone power is shown as a function of frequency with (blue) and without (yellow) the FC pump turned on. Here, it is calculated as the ratio of the transmission when the dc bias and the pumps are turned on, to the transmission when the dc bias and the pumps are turned off. (b) The TWPAC-input $1$\,dB compression point is shown with (blue) and without (yellow) the FC pump turned on. The cut portion around $8.8$\,GHz corresponds to a situation where $P_{1\mathrm{dB}}$ was not yet reached at the highest VNA probe tone power.}
    \label{fig:1dbcp}
\end{figure*}

Figure \ref{fig:1dbcp} shows the input $1$\,dB compression point, $P_{1\mathrm{dB}}$, and the corresponding small signal TWPAC gain, here defined as the ratio of the VNA transmission when the dc bias and the pumps are turned on, to the transmission when the dc bias and the pumps are turned off. The $P_{1\mathrm{dB}}$ is the point at which the gain drops by $1$\,dB when increasing the VNA probe power. Between $5.5$ and $9$\,GHz, $P_{1\mathrm{dB}}\approx-90$\,dBm on average, with a corresponding average TWPAC gain $G\approx7.3$\,dB. Assuming that $P_{1\mathrm{dB}}$ will decrease with increasing gain, it is on par with conventional Josephson-based TWPAs \cite{macklin2015near,planat2020photonic}, for which $P_{1\mathrm{dB}}$ is $\approx-100$\,dBm at $\approx20$\,dB gain. Note also that when the FC pump is turned off, $P_{1\mathrm{dB}}$ stays at the same level, whereas the gain increases to $\approx7.9$\,dB on average between $5.5$ and $9$\,GHz, indicating that for the same gain, turning on the FC marginally diminishes the output compression point by $\approx0.6$\,dB.

The input attenuation of the line, from the VNA (emitting the probe tone) to the TWPAC input is calibrated as follows: we first perform a noise measurement using the SNTJ, with the TWPAC off (pumps off, no dc bias), from which we obtain the system-added noise and the gain of the chain $G_c^\mathrm{off}$, from the SNTJ's reference plane to the spectrum analyzer (SA). We then measure the output power $P_\mathrm{VNA}^\mathrm{out}$ on the SA when sending a VNA probe tone, varying its frequency (TWPAC still off). Finally, we divide the probe tone power emitted by the VNA $P_\mathrm{VNA}^\mathrm{in}$ by the tone power, referred to the SNTJ's reference plane $P_\mathrm{VNA}^\mathrm{out}/G_c^\mathrm{off}$, which gives us a good estimate of the line's input attenuation $A$, from the VNA emitting port to the TWPAC input:
\begin{equation}
    A = \frac{P_\mathrm{VNA}^\mathrm{in}}{P_\mathrm{VNA}^\mathrm{out}/G_c^\mathrm{off}}.
\end{equation}
Note that this attenuation includes the loss between the SNTJ and the TWPAC, as such it is an over-estimate of the attenuation between the VNA emitting port and the TWPAC chip-input. As a consequence, the input $1$\,dB compression point is slightly under-estimated.

\subsubsection{The TWPAC output spectrum}
\label{sec:TWPACoutspect}

\begin{figure*}[h]
    \vspace{1pt}
	\includegraphics[scale=0.95]{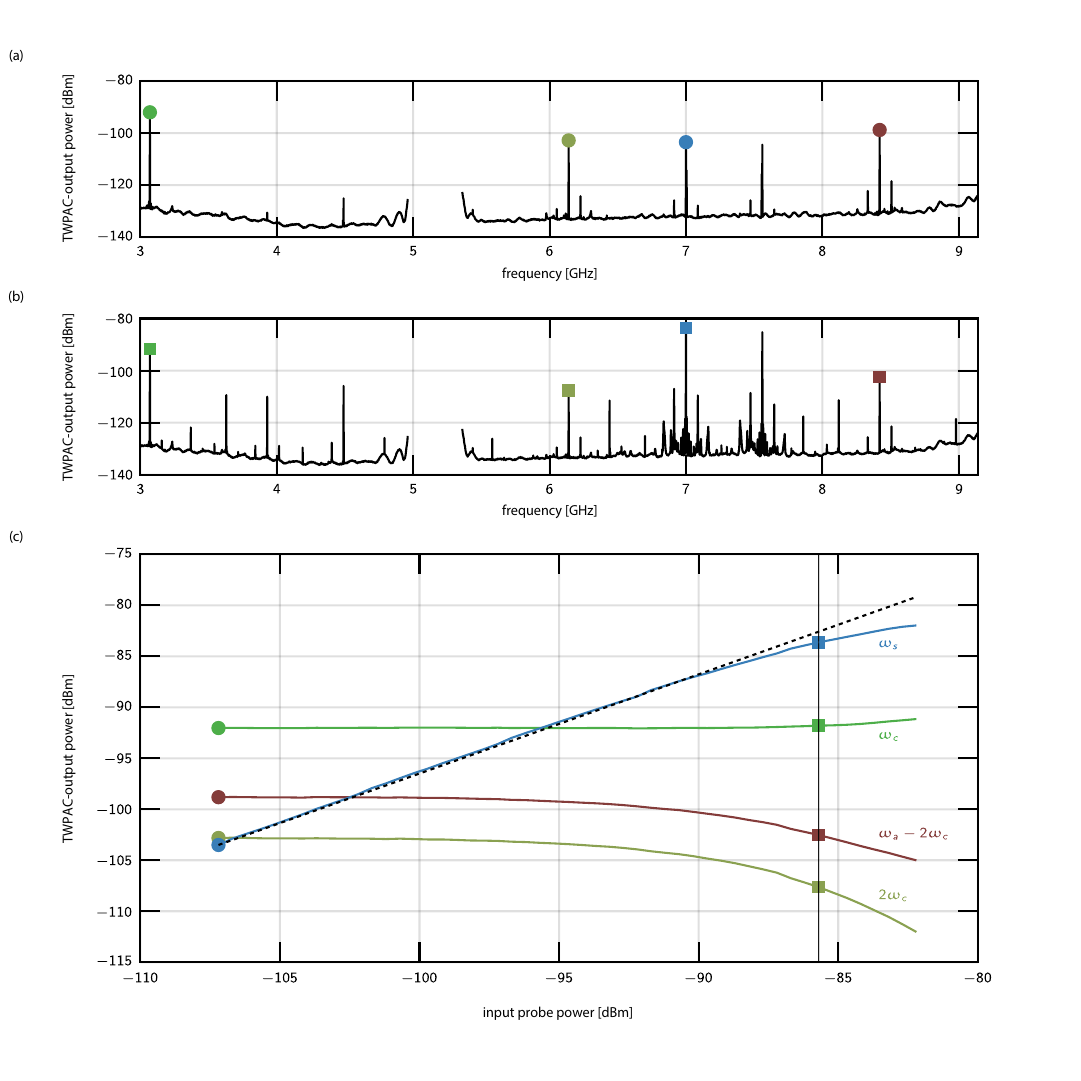}
    \caption{The evolution of the TWPAC output spectrum, measured on the SA, when increasing a CW probe tone at $7$\,GHz. (a) The output spectrum at low probe tone power presents $5$ major peaks at $\omega_c$, $\omega_{c_2}=2\omega_c$, $\omega_s$, $\omega_i=\omega_a-\omega_s$, and $\omega_a-2\omega_c$. We cut the spectrum at the first TWPAC stopband, because within the stopband the gain of the chain cannot be accurately measured, therefore the power of any bin within the stopband cannot be accurately compared to the other bins. (b) The output spectrum when sending a probe tone power $P_{1\mathrm{dB}}=-86$\,dBm contains more peaks, that correspond to activating more parametric processes. (c) The output power of the few selected modes, referred to the TWPAC output is plotted as a function of the input probe tone power, referred to the TWPAC input. The vertical line indicates $P_{1\mathrm{dB}}$ for the CW tone at $\omega_s=7$\,GHz. The input probe tone power calibration was the same as in the Methods, sec.\,\ref{sec:P1db}, and we refer the output power to the TWPAC output by dividing the power measured on the SA by the chain's gain, measured with the SNTJ, with the TWPAC off.}
    \label{fig:compression_tones}
\end{figure*}

Experimentally, the FC pump frequency $\omega_c$ being lower than what was expected from the original design, its second harmonic $\omega_{c_2}=2\omega_c$ lies within the amplification bandwidth, together with its corresponding idler $\omega_a-2\omega_c$, where $\omega_a$ is the PA pump frequency. One can wonder whether these modes limit the TWPAC power handling. In Fig.\,\ref{fig:compression_tones}a and \ref{fig:compression_tones}b we show the TWPAC output spectrum measured on the SA, when sending a probe tone at $\omega_s/2\pi = 7$\,GHz, at low power (a) and at the power $P_{1\mathrm{dB}}$ for that frequency (b). We identify the modes of interest at $\omega_c$, $\omega_s$, $2\omega_c$, and $\omega_a-2\omega_c$. Evidently, there are some other peaks, in particular we notice side-bands on each of these main modes, detuned by $\pm87$\,MHz, which might come from noise in the dc biasing line of the TWPAC. At high tone power (Fig.\,\ref{fig:compression_tones}b), more peaks arise from the spectrum noise floor compared to the low power case, because more parametric processes become non-negligible, and the input probe tone power get scattered onto various modes. This is what saturates the amplifier.

In Fig.\,\ref{fig:compression_tones}c we plot the output power of the few selected modes of interest, referred to the TWPAC output, as a function of the $7$\,GHz input probe tone power (referred to the TWPAC input). Taken individually, none of these tones seem to be predominantly driving the TWPAC saturation, because their power remain much below the TWPAC saturation power $P_{1\mathrm{dB}}$ at that particular frequency. Similarly, we see in Fig.\,\ref{fig:compression_tones}b that there exists parasitic 4WM processes around all the tones and pump harmonics when reaching $P_\mathrm{1dB}$, but these are even lower in power than the $c_2$ amplified tone and idler, therefore none of these tones should contribute to limiting the gain. In the next section we show that the parasitic 3WM amplification of $c_2$ and parasitic 4WM processes do not alter the noise performance, hinting at the fact that they likely do not contribute predominantly to reducing the gain.

\subsection{Effect of the frequency conversion on the gain and on noise measurements}
\label{sec:FCongainandnoise}

In Fig.\,\ref{fig:noise} we presented the noise measurement when the TWPAC is driven with both the PA and FC pumps. Here we show that the FC pump does not degrade the TWPAC noise performance, nor does it seem to saturate the amplifier.

\begin{figure*}[h]
    \vspace{1pt}
	\includegraphics[scale=0.95]{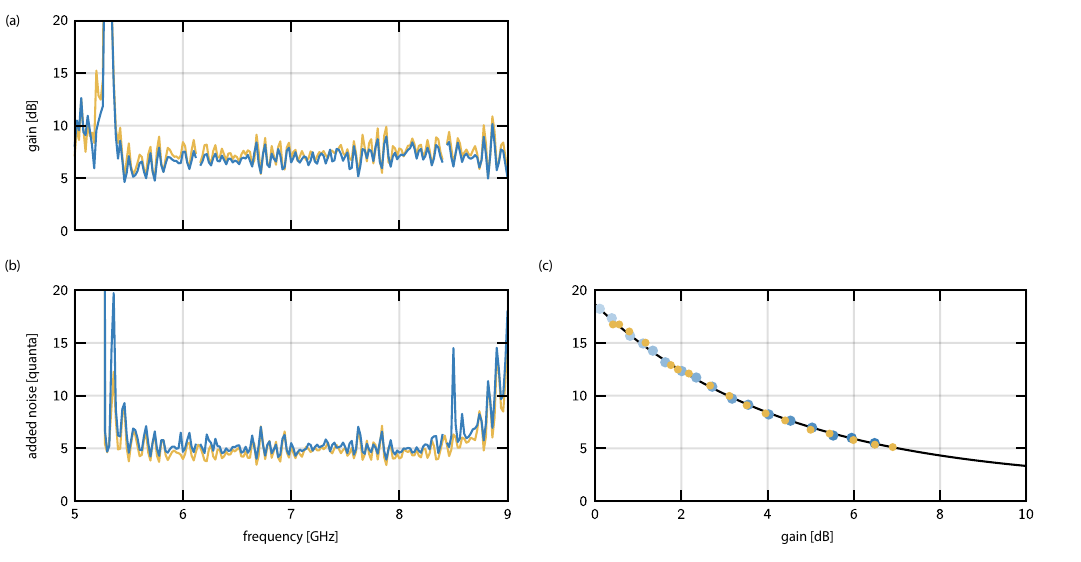}
    \caption{Effect of the FC pump on the gain and added noise. (a) The on/off gain (\ie, not referenced to a through) as a function of frequency, when the FC pump is off (yellow) and when it is on (blue). (b) The added noise is plotted within the same frequency range, when the FC pump is off (yellow) and when it is on (blue). (c) The added noise of the bin centered at $7$\,GHz is plotted as a function of the TWPAC gain, when the FC pump is off (yellow) and when it is on (blue). The black line represents the model $N_\mathrm{sys}=N_1+N_2/G$.}
    \label{fig:comparison_noise}
\end{figure*}

Figure \ref{fig:comparison_noise}a shows the on/off gain (\ie{} not referenced to a through) as a function of frequency, measured when performing the noise characterization experiment, with only the PA pump turned on, and when both the PA and FC pumps are turned on. We can notice (i) that the amount of gain is comparable to that presented in Fig.\,\ref{fig:scatdata}b, which suggests that the TWPAC is not very sensitive to the characterization setup it is embedded in (Fig.\,\ref{fig:scatdata}a or Fig.\,\ref{fig:noise}a). (ii) As expected, the on/off gain is flatter compared to that referenced to a through (see Fig.\,\ref{fig:scatdata}a). (iii) The gain slightly lowers when the FC pump is turned on, but it should not be interpreted as gain compression. In fact, Fig.\,\ref{fig:comparison_noise}b shows the measured TWPAC noise as a function of frequency, when the FC is on and off. On average, when the FC is off the noise between $5.5$ and $8.5$\,GHz is $4.8$ quanta, slightly lower than when the FC is on ($5.2$ quanta). However, this discrepancy is perfectly explained by the slight gain variation existing between the two situations with FC pump on and FC off (in both situations the PA pump is kept at the same power and frequency). When plotting the added noise $N_\mathrm{add}$ as a function of the TWPAC gain for the frequency bin centered at $7$\,GHz (Fig.\,\ref{fig:comparison_noise}c), both situations (FC on and FC off) overlap on the same curve describing the model $N_\mathrm{add}=N_1+N_2/G$ (where $N_1$ is the noise added by the TWPAC and by all the lossy elements preceding it, up to the SNTJ, and where $N_2$ is the noise added by the elements in the chain after the TWPAC). Therefore, the intrinsic added noise of the TWPAC stays the same, regardless of having the FC pump turned on or off, and the parasitic 3WM PA of $c_2$ and 4WM around the FC pumps harmonics do not seem to be the main reason for the limited gain we experimentally obtained.

\section*{Data Availability}

The data that support the findings of this study are available from the NIST Public Data Repository \cite{nistdata} at https://doi.org/10.18434/mds2-3875. Source data are provided with this paper.

\section*{Code Availability}

The codes used for the simulations are available from the corresponding authors on reasonable request.

\section*{Acknowledgements}

We thank Ryan Kaufman, Connor Denney, Arpit Ranadive, Nicolas Roch, Matthieu Praquin, and Philippe Campagne-Ibarcq for fruitful discussions. We thank Manuel Castellanos-Beltran and Cody Scarborough for their feedback on the manuscript. Certain commercial materials and equipment are identified in this paper to foster understanding. Such identification does not imply recommendation or endorsement by the National Institute of Standards and Technology, nor does it imply that the materials or equipment identified is necessarily the best available for the purpose.

\section*{Author Contributions}

M.M., J.D.T., J.A. and F.L. conceived and designed the experiment. M.M. designed and simulated the device. B.T.M. and J.A.E. designed the packaging. M.M. performed the measurements. M.M. and F.L. analyzed the data. K.G. and K.C. fabricated the device. M.M., B.T.M., J.D.T., J.A. and F.L. wrote the manuscript.

\section*{Competing Interests}

The authors declare no competing interests.

\clearpage

%


\clearpage
\widetext
\begin{center}
\textbf{\large Supplementary information: A Travelling Wave Parametric Amplifier and Converter}
\end{center}
\setcounter{equation}{0}
\setcounter{figure}{0}
\setcounter{table}{0}
\makeatletter
\renewcommand{\theequation}{S\arabic{equation}}
\renewcommand{\thefigure}{S\arabic{figure}}
\renewcommand{\thetable}{S\arabic{table}}
\renewcommand{\theHequation}{S\arabic{equation}}
\renewcommand{\theHfigure}{S\arabic{figure}}
\renewcommand{\theHtable}{S\arabic{table}}
\renewcommand{\bibnumfmt}[1]{[S#1]}
\renewcommand{\citenumfont}[1]{S#1}

\section{The TWPAC coupled-mode equations}

Here we derive the equations of motion governing the propagation of the few modes of interest through the TWPAC: amplification pump (frequency $\omega_a$), signal ($\omega_s$), idler ($\omega_i)$, conversion pump ($\omega_c$), down-converted signal ($\omega_d$), up-converted signal ($\omega_u$) and second harmonic of the conversion pump ($\omega_{c_2}$). Our model includes the reflections of these tones at the TWPAC's port, due to the imperfect impedance matching between the TWPAC's (frequency-dependent) impedance and its environment, closely following the formalism of Kern. et al. \cite{Kern2023reflection_s}.

\subsection{Equivalent inductance of a dc-biased Josephson junction}

We start by deriving the equivalent nonlinear inductance to a dc-biased Josephson junction (JJ). A JJ's equivalent inductance is:
\begin{equation}
    L_J = \frac{L_{J0}}{\sqrt{1-\left(\frac{I_\Sigma}{I_c}\right)^2}},
\end{equation}
where $L_{J0}=\phi_0/I_c$ is the JJ's unbiased inductance, with $\phi_0$ the reduced flux quantum, with $I_c$ the JJ's critical current, and where $I_\Sigma$ is the total current traversing the JJ. If we use a fixed dc bias $I_d$ such that $I_\Sigma = I_d + I$, with $I_\Sigma\ll I_c$ is a time-varying oscillating term, we can expand the inductance around the dc working point:
\begin{equation}
    L_J = L_d[1 + \epsilon I + \xi I^2 + \mathcal{O}(I^3)],
\end{equation}
with
\begin{align}
    L_d &= \frac{L_{J0}}{\sqrt{1-\left(\frac{I_d}{I_c}\right)^2}} \\
    \epsilon &= \frac{I_d}{I_c^2-I_d^2} \\
    \xi &= \frac{I_c^2+2I_d^2}{2(I_c^2-I_d^2)^2}
\end{align}
In practice, the JJ is at the interface between two metallic surfaces, which therefore add a capacitance $C_J$, in parallel to this effective nonlinear inductance. Below its resonant frequency (called the plasma frequency $\omega_p = 1/\sqrt{C_JL_d}$), the tank circuit is equivalent to an effective inductance $\Tilde{L}_J$ such that:
\begin{equation}
    \Tilde{L}_J = \frac{L_J}{1-C_JL_J\omega^2} = \Tilde{L}_d[1+\Tilde{\epsilon}I + \Tilde{\xi}I^2 + \mathcal{O}(I^3)],
    \label{eq:Ljtilde}
\end{equation}
where
\begin{align}
    \Tilde{L}_d &= \frac{L_d}{1-\left(\frac{\omega}{\omega_p}\right)^2}  \label{eq:tildeLd}\\
    \Tilde{\epsilon} &= \frac{\epsilon}{1-\left(\frac{\omega}{\omega_p}\right)^2} \\
    \Tilde{\xi} &= \frac{\xi+(\epsilon^2-\xi)\left(\frac{\omega}{\omega_p}\right)^2}{\left[1-\left(\frac{\omega}{\omega_p}\right)^2\right]^2}.
\end{align}
In the following, we drop the $\Tilde{}$ notation, but will remember that $L_d$, $\epsilon$ and $\xi$ now all depend on $\omega$, although far from the plasma frequency, this dependence is negligible. In our case $\omega_p/2\pi\approx40$\,GHz. When ambiguous, we will explicitly write this dependence.

\subsection{The coupled-mode equations}
\label{sec:CME}

Writing the nonlinear inductance up to the second order in $I$ (Eq.\,\ref{eq:Ljtilde}) allows us to include the 3WM-type of nonlinearity (which comes from the term proportional to $\epsilon$), as well as gain compression coming from the 4WM (\ie{} Kerr) term proportional to $\xi$. In the first, simple model presented in Fig.\,\ref{fig:design}, we did not include the effect of the 4WM term (equivalent to taking $\xi=0$ in the following equations). We then included them when comparing the measurements to the CME, see Sec.\,\ref{sec:upconv}.

Writing the the telegraphers' equations in this context, including both 3WM and 4WM, we obtain the propagation equation \cite{Malnou2021three_s}
\begin{equation}
    v_p^2 \frac{\partial^2I}{\partial x^2} - \frac{\partial^2 I}{\partial t^2} = \frac{\partial^2}{\partial t^2}\left(\frac{\epsilon}{2} I^2 + \frac{\xi}{3}I^3\right),
\label{eq:telegrapher}
\end{equation}
where $v_p = 1/\sqrt{L_d(\omega)C}$ is the line's phase velocity, $x$ is the position along the transmission line (which for us will be in units of cells) and $t$ is the time.

We now perform a harmonic balance, \ie{} we assume that the current $I(x,t)$, solution of the equation of motion Eq.\,\ref{eq:telegrapher}, is the sum of the selected propagating modes of interest, indexed by $n\in\{a,s,i,c,d,u,c_2\}$. The modes propagate forward or backward, and we study two situations, depending on whether we look at the forward or backward response of the device. For the forward response, we assume that $\{a,s,i,d,u,c_2\}$ are traveling forward, whereas $c$ travels backward, and for the backward response, we assume that $\{s,i,c,d,u,c_2\}$ are traveling backward, while $a$ travels forward, which by symmetry is equivalent to considering $\{s,i,c,d,u,c_2\}$ as traveling forward, while $a$ travels backward. Table \ref{tab:modesprop} summarizes the way the modes are propagating. Note that (i) we account for possible reflections of all these modes at the TWPAC ports following \cite{Kern2023reflection_s}, and (ii) that formally, the only thing that differentiates a mode traveling forward to a mode traveling backward is that for the backward traveling mode, we need to do the transformation $k_n\leftarrow-k_n$, where $k_n$ is the mode wavenumber, in the coupled-mode equations. 
\begin{table}[htbp]
        \centering
        \begin{tabular}{l|ll}
         & modes traveling forward & modes traveling backward \\
        \hline
        forward response & $a,s,i,d,u,c_2$ & $c$ \\
        backward response & $s,i,d,u,c,c_2$ & $a$ \\
        \end{tabular}
        \caption{List and propagation orientation of the modes, when looking at the forward an backward response of the TWPAC.}
        \label{tab:modesprop}
\end{table}

Thus the current, solution of Eq.\,\ref{eq:telegrapher} is
\begin{equation}    
\begin{aligned}
    I(x,t) &= \sum_{n\in\{a,s,i,c,d,u,c_2\}}\left\{\frac{1}{2}I_n(x)\left[e^{\pm ik_nx} + \Gamma_n e^{\pm ik_nN}e^{\mp ik_nx} + \left(\Gamma_n e^{\pm ik_nN}\right)^2e^{\pm ik_nx} + ... \right]e^{-i\omega_n t} + \mathrm{c.c.}\right\}\\
    &= \sum_{n\in\{a,s,i,c,d,u,c_2\}}\left[\frac{1}{2}I_n(x)t_n\left(e^{\pm ik_nx} + \Tilde{\Gamma}_n e^{\mp ik_nx}\right)e^{-i\omega_n t} + \mathrm{c.c.}\right] \\
    &= \sum_{n\in\{a,s,i,c,d,u,c_2\}} \frac{1}{2}\alpha_n^\pm + \frac{1}{2}\alpha_n^{*\pm},
    \label{eq:I}
\end{aligned}
\end{equation}
where 
\begin{equation}
    \alpha_n^\pm = I_n(x)t_n\left(e^{\pm ik_nx} + \Tilde{\Gamma}_n e^{\mp ik_nx}\right)e^{-i\omega_n t},
\end{equation}
with the $\pm$ denoting the forward ($+$) or backward ($-$) traveling modes. For clarity, in what follows we omit the superscript $+$ for the modes $\{s,i,d,u,c_2\}$, because they always travel forward (see Tab.\,\ref{tab:modesprop}). Here, c.c. is the complex conjugate, $I_n(x)$ is the complex field amplitude of mode $n$, and
\begin{align}
    t_n = \frac{1}{1-\Gamma_n e^{i2k_nN}}\\
    \Tilde{\Gamma}_n = \Gamma_n e^{ik_nN},
\end{align}
with $N$ the total length of the TWPAC, in units of cell-lengths (so $N$ is the total number of cells), and $\Gamma_n$ is the amplitude reflection coefficient of mode $n$. In other words, if the TWPAC impedance at $\omega_n$ is $Z_n$, and if the TWPAC is embedded in an environment with impedance $Z_0$, we have
\begin{equation}
    \Gamma_n = \frac{\lvert Z_n-Z_0 \rvert}{Z_n+Z_0}.
\end{equation}

We now need to replace $I$ in Eq.\,\ref{eq:telegrapher} with $I(x,t)$ from Eq.\,\ref{eq:I}, and then collect the terms at each modes of interest. First, we deal with the left-hand-side (LHS) of Eq.\,\ref{eq:telegrapher}:
\begin{equation}
    \mathrm{LHS} = v_p^2\sum_{n\in\{a,s,i,c,d,u,c_2\}}\left[\pm ik_nt_n \frac{\partial I_n(x)}{\partial x}\left(e^{\pm ik_nx}-\Tilde{\Gamma}_n e^{\mp ik_nx}\right)e^{-i\omega_nt} + \mathrm{c.c.}\right].
\end{equation}
Therefore, for any mode $n$ we obtain:
\begin{equation}
    \mathrm{LHS}\Bigr|_{\substack{\omega=\omega_n}} = \pm i \frac{\omega_n^2}{k_n}t_n \frac{\partial I_n(x)}{\partial x}\left(e^{\pm ik_nx}-\Tilde{\Gamma}_n e^{\mp ik_nx}\right)e^{-i\omega_nt} + \mathrm{c.c.},
\end{equation}
using the fact that $v_p=\omega_n/k_n$.

On the right-hand-side (RHS), we perform a rotating wave approximation: we only keep the terms whose frequencies are at one of the modes $n\in\{a,s,i,c,d,u,c_2\}$. For the three-wave mixing processes, coming from the term $\partial^2 I^2/\partial t^2$ in Eq.\,\ref{eq:telegrapher}, we obtain the coefficients gathered in Tab.\,\ref{tab:3WMproc}. For the four-wave mixing processes, coming from the term $\partial^2 I^3/\partial t^2$ in Eq.\,\ref{eq:telegrapher}, we obtain the coefficients gathered in Tab.\,\ref{tab:4WMproc}. Note that for both 3WM and 4WM, all of the permutations of a given process have to be accounted for.


\begin{table}[htbp]
        \centering
        \begin{tabular}{lll}
        \hline
        \hline        
        3WM process & coefficient \\
        \hline        
        $\omega_s+\omega_i=\omega_a$ & $-\omega_a^2\frac{1}{2}\alpha_s\alpha_i$ \\
        $\omega_a-\omega_i=\omega_s$ & $-\omega_s^2\frac{1}{2}\alpha_a^\pm\alpha_i^*$ \\
        $\omega_a-\omega_s=\omega_i$ & $-\omega_i^2\frac{1}{2}\alpha_a^\pm\alpha_s^*$ \\        
        \hline
        $\omega_c+\omega_d=\omega_s$ & $-\omega_s^2\frac{1}{2}\alpha_c^\mp\alpha_d$ \\
        $\omega_s-\omega_d=\omega_c$ & $-\omega_c^2\frac{1}{2}\alpha_s\alpha_d^*$ \\
        $\omega_s-\omega_c=\omega_d$ & $-\omega_d^2\frac{1}{2}\alpha_s\alpha_c^{*\mp}$ \\
        \hline
        $\omega_u-\omega_c=\omega_s$ & $-\omega_s^2\frac{1}{2}\alpha_u\alpha_c^{*\mp}$ \\
        $\omega_u-\omega_s=\omega_c$ & $-\omega_c^2\frac{1}{2}\alpha_u\alpha_s^*$ \\
        $\omega_s+\omega_c=\omega_u$ & $-\omega_u^2\frac{1}{2}\alpha_s\alpha_c^\mp$ \\
        \hline
        $\omega_c+\omega_c=\omega_{c_2}$ & $-\omega_{c_2}^2\frac{1}{4}\alpha_c^\mp\alpha_c^\mp$ \\
        $\omega_{c_2}-\omega_c=\omega_c$ & $-\omega_c^2\frac{1}{2}\alpha_{c_2}\alpha_c^{*\mp}$ \\
        \hline
        $\omega_u-\omega_d=\omega_{c_2}$ & $-\omega_{c_2}^2\frac{1}{2}\alpha_u\alpha_d^*$ \\
        $\omega_u-\omega_{c_2}=\omega_d$ & $-\omega_d^2\frac{1}{2}\alpha_u\alpha_{c_2}^*$ \\
        $\omega_{c_2}+\omega_d=\omega_u$ & $-\omega_u^2\frac{1}{2}\alpha_{c_2}\alpha_d$ \\
        \hline
        \hline
        \end{tabular} 
        \caption{List of all the 3WM processes and their coefficients when restricting to the mode basis $\{a,s,i,c,d,u,c_2\}$. For clarity, each permutations of a given process are grouped together. For the PA pump (mode $a$) and FC pump (mode $c$), we distinguish the forward ($+$) and backward ($-$) traveling modes.}
        \label{tab:3WMproc}
\end{table}
\begin{table}[h!] 
        \centering
        \begin{tabular}{lll}
        \hline
        \hline        
        4WM process & coefficient \\
        \hline        
        $\omega_a+\omega_a-\omega_a=\omega_a$ & $-\omega_a^2\frac{3}{8}\alpha_a^\pm\lvert \alpha_a^\pm\rvert^2$ \\
        $\omega_a+\omega_c-\omega_c=\omega_a$ & $-\omega_a^2\frac{6}{8}\alpha_a^\pm\lvert \alpha_c^\mp\rvert^2$ \\
        $\forall n\in\{s,i,d,u,c_2\},\hspace{0.1cm} \omega_a+\omega_n-\omega_n=\omega_a$ & $-\omega_a^2\frac{6}{8}\alpha_a^\pm\lvert \alpha_n\rvert^2$ \\            
        \hline
        $\omega_c+\omega_c-\omega_c=\omega_c$ & $-\omega_c^2\frac{3}{8}\alpha_c^\mp\lvert \alpha_c^\mp\rvert^2$ \\
        $\omega_c+\omega_a-\omega_a=\omega_c$ & $-\omega_c^2\frac{6}{8}\alpha_c^\mp\lvert \alpha_a^\pm\rvert^2$ \\
        $\forall n\in\{s,i,d,u,c_2\},\hspace{0.1cm} \omega_c+\omega_n-\omega_n=\omega_c$ & $-\omega_c^2\frac{6}{8}\alpha_c^\mp\lvert \alpha_n\rvert^2$ \\            
        \hline
        \multicolumn{2}{c}{$\forall m \in \{s,i,d,u,c_2\}$}\\        
        $\omega_m+\omega_m-\omega_m=\omega_m$ & $-\omega_m^2\frac{3}{8}\alpha_m\lvert \alpha_m\rvert^2$ \\
        $\omega_m+\omega_a-\omega_a=\omega_m$ & $-\omega_m^2\frac{6}{8}\alpha_m\lvert \alpha_a^\pm\rvert^2$ \\
        $\omega_m+\omega_c-\omega_c=\omega_m$ & $-\omega_m^2\frac{6}{8}\alpha_m\lvert \alpha_c^\mp\rvert^2$ \\
        $\forall n\in\{s,i,d,u,c_2\}\hspace{0.1cm}\&\hspace{0.1cm} n\neq m,\hspace{0.1cm} \omega_m+\omega_n-\omega_n=\omega_m$ & $-\omega_m^2\frac{6}{8}\alpha_m\lvert \alpha_n\rvert^2$ \\             
        \hline        
        $\omega_u-\omega_d-\omega_c=\omega_c$ & $-\omega_c^2\frac{6}{8}\alpha_u\alpha_d^*\alpha_c^{*\mp}$ \\        
        $\omega_u-\omega_c-\omega_c=\omega_d$ & $-\omega_d^2\frac{3}{8}\alpha_u(\alpha_c^{*\mp})^2$ \\        
        $\omega_d+\omega_c+\omega_c=\omega_u$ & $-\omega_u^2\frac{3}{8}\alpha_d(\alpha_c^{\mp})^2$ \\        
        \hline                
        $\omega_c+\omega_s-\omega_d=\omega_{c_2}$ & $-\omega_{c_2}^2\frac{6}{8}\alpha_c^\mp\alpha_s\alpha_d^*$ \\                
        $\omega_c+\omega_s-\omega_{c_2}=\omega_d$ & $-\omega_d^2\frac{6}{8}\alpha_c^\mp\alpha_s\alpha_{c_2}^*$ \\                
        $\omega_{c_2}-\omega_c+\omega_d=\omega_s$ & $-\omega_s^2\frac{6}{8}\alpha_{c_2}\alpha_{c}^{*\mp}\alpha_d$ \\                
        $\omega_{c_2}-\omega_s+\omega_d=\omega_c$ & $-\omega_c^2\frac{6}{8}\alpha_{c_2}\alpha_s^*\alpha_d$ \\                
        \hline                
        $\omega_u-\omega_s+\omega_c=\omega_{c_2}$ & $-\omega_{c_2}^2\frac{6}{8}\alpha_u\alpha_s^*\alpha_c^{\mp}$ \\                        
        $\omega_u-\omega_{c_2}+\omega_c=\omega_s$ & $-\omega_s^2\frac{6}{8}\alpha_u\alpha_{c_2}^*\alpha_c^{\mp}$ \\                        
        $\omega_{c_2}-\omega_c+\omega_s=\omega_u$ & $-\omega_u^2\frac{6}{8}\alpha_{c_2}\alpha_c^{*\mp}\alpha_s$ \\                        
        $\omega_{c_2}-\omega_u+\omega_s=\omega_c$ & $-\omega_c^2\frac{6}{8}\alpha_{c_2}\alpha_u^*\alpha_s$ \\        
        \hline                
        $\omega_a-\omega_i+\omega_c=\omega_u$ & $-\omega_u^2\frac{6}{8}\alpha_a^\pm\alpha_i^*\alpha_c^{\mp}$ \\
        $\omega_a-\omega_u+\omega_c=\omega_i$ & $-\omega_i^2\frac{6}{8}\alpha_a^\pm\alpha_u^*\alpha_c^{\mp}$ \\
        $\omega_u-\omega_a+\omega_i=\omega_c$ & $-\omega_c^2\frac{6}{8}\alpha_u\alpha_a^{*\pm}\alpha_i$ \\
        $\omega_u-\omega_c+\omega_i=\omega_a$ & $-\omega_a^2\frac{6}{8}\alpha_u\alpha_c^{*\mp}\alpha_i$ \\
        \hline                
        $\omega_a-\omega_i-\omega_c=\omega_d$ & $-\omega_d^2\frac{6}{8}\alpha_a^\pm\alpha_i^*\alpha_c^{*\mp}$ \\
        $\omega_a-\omega_d-\omega_c=\omega_i$ & $-\omega_i^2\frac{6}{8}\alpha_a^\pm\alpha_d^*\alpha_c^{*\mp}$ \\
        $\omega_a-\omega_d-\omega_i=\omega_c$ & $-\omega_c^2\frac{6}{8}\alpha_a^\pm\alpha_d^*\alpha_i^*$ \\
        $\omega_d+\omega_i+\omega_c=\omega_a$ & $-\omega_a^2\frac{6}{8}\alpha_d\alpha_i\alpha_c^\mp$ \\
        \hline
        \hline
        \end{tabular} 
        \caption{List of all the 4WM processes and their coefficients when restricting to the mode basis $\{a,s,i,c,d,u,c_2\}$. For clarity, family of processes are grouped together.}
        \label{tab:4WMproc}
\end{table}

\clearpage
Finally, we join the LHS and RHS together, to get the coupled-mode equations (and their complex conjugate, not shown):
\begin{align}
&\begin{aligned}
\frac{\partial I_a}{\partial x} = \pm &i \frac{\epsilon(\omega_a)}{4} \frac{k_a}{t_a} \mathcal{F}^{si}_{\pm a} t_st_iI_sI_i \\
\pm &i \frac{\xi(\omega_a)}{8}k_aI_a\left(\mathcal{F}^{\pm a \pm a \pm a^*}_{\pm a}\lvert t_a\rvert^2 \lvert I_a\rvert^2  + 2 \mathcal{F}^{\pm a \mp c \mp c^*}_{\pm a}\lvert t_c\rvert^2\lvert I_c \rvert^2 + \sum_{m\in\{s,i,d,u,c_2\}}2\mathcal{F}^{\pm a m m^*}_{ \pm a}\lvert t_m\rvert^2\lvert I_m \rvert^2\right) \\
\pm &i \frac{\xi(\omega_a)}{8}\frac{k_a}{t_a}\left( 2\mathcal{F}^{u\mp c^* i}_a t_u t_c^* t_i I_u I_c^* I_i + 2\mathcal{F}^{d i \mp c}_a t_d t_i t_c I_d I_i I_c\right)
    \label{eq:CMEa}
\end{aligned}\\
&\begin{aligned}
\frac{\partial I_s}{\partial x} = &i \frac{\epsilon(\omega_s)}{4} \frac{k_s}{t_s} \left( \mathcal{F}^{\pm ai^*}_s t_at_i^* I_aI_i^* + \mathcal{F}^{\mp cd}_s t_ct_dI_cI_d + \mathcal{F}^{u\mp c^*}_s t_ut_c^*I_uI_c^*\right) \\
+ &i \frac{\xi(\omega_s)}{8}k_sI_s\left(\mathcal{F}^{sss^*}_s\lvert t_s\rvert^2 \lvert I_s\rvert^2 + 2\mathcal{F}^{s\pm a \pm a^*}_s\lvert t_a\rvert^2\lvert I_a \rvert^2 + 2\mathcal{F}^{s\mp c \mp c^*}_s\lvert t_c\rvert^2\lvert I_c \rvert^2 + \sum_{m\in\{i,d,u,c_2\}}2\mathcal{F}^{smm^*}_s\lvert t_m\rvert^2\lvert I_m \rvert^2\right) \\
+ &i \frac{\xi(\omega_s)}{8}\frac{k_s}{t_s}\left( 2\mathcal{F}^{c_2 \mp c^* d}_s t_{c_2} t_c^* t_d I_{c_2} I_c^* I_d + 2\mathcal{F}^{u c_2^* \mp c}_s t_u t_{c_2}^* t_c I_u I_{c_2}^* I_c\right)
\label{eq:CMEs}
\end{aligned}\\
&\begin{aligned}
\frac{\partial I_i}{\partial x} = &i \frac{\epsilon(\omega_i)}{4} \frac{k_i}{t_i} \mathcal{F}^{\pm as^*}_i t_at_s^* I_aI_s^* \\
+ &i \frac{\xi(\omega_i)}{8}k_iI_i\left(\mathcal{F}^{iii^*}_i\lvert t_i\rvert^2 \lvert I_i\rvert^2 + 2\mathcal{F}^{i\pm a \pm a^*}_i\lvert t_a\rvert^2\lvert I_a \rvert^2 + 2\mathcal{F}^{i\mp c \mp c^*}_i\lvert t_c\rvert^2\lvert I_c \rvert^2 + \sum_{m\in\{s,d,u,c_2\}}2\mathcal{F}^{imm^*}_i\lvert t_m\rvert^2\lvert I_m \rvert^2\right) \\
+ &i \frac{\xi(\omega_i)}{8}\frac{k_i}{t_i}\left( 2\mathcal{F}^{\pm a u^* \mp c}_i t_a t_u^* t_c I_a I_u^* I_c + 2\mathcal{F}^{\pm a d^* \mp c^*}_i t_a t_d^* t_c^* I_a I_d^* I_c^*\right)
\label{eq:CMEi}
\end{aligned}\\
&\begin{aligned}
\frac{\partial I_c}{\partial x} = \mp &i \frac{\epsilon(\omega_c)}{4} \frac{k_c}{t_c} \left(\mathcal{F}^{sd^*}_{\mp c} t_st_d^* I_sI_d^* + \mathcal{F}^{us^*}_{\mp c} t_ut_s^*I_uI_s^* + \mathcal{F}^{c_2\mp c^*}_{\mp c} t_{c_2}t_c^*I_{c_2}I_c^*\right) \\
\mp &i \frac{\xi(\omega_c)}{8}k_cI_c\left(\mathcal{F}^{\mp c \mp c \mp c^*}_{\mp c}\lvert t_c\rvert^2 \lvert I_c\rvert^2 + 2\mathcal{F}^{\mp c\pm a \pm a^*}_{\mp c}\lvert t_a\rvert^2\lvert I_a \rvert^2 + \sum_{m\in\{s,i,d,u,c_2\}}2\mathcal{F}^{\mp cmm^*}_{\mp c}\lvert t_m\rvert^2\lvert I_m \rvert^2\right)\\
\mp &i \frac{\xi(\omega_c)}{8}\frac{k_c}{t_c}\Big(2\mathcal{F}^{u d^* \mp c^*}_{\mp c} t_u t_d^* t_c^* I_u I_d^* I_c^* + 2\mathcal{F}^{c_2 s^* d}_{\mp c} t_{c_2} t_s^* t_d I_{c_2} I_s^* I_d + 2\mathcal{F}^{c_2 u^* s}_{\mp c} t_{c_2} t_u^* t_s I_{c_2} I_u^* I_s\\ 
+ &2\mathcal{F}^{u \pm a^* i}_{\mp c} t_u t_a^* t_i I_u I_a^* I_i + 2\mathcal{F}^{\pm a d^* i^*}_{\mp c} t_a t_d^* t_i^* I_a I_d^* I_i^*\Big)
\label{eq:CMEc}
\end{aligned}\\
&\begin{aligned}
\frac{\partial I_d}{\partial x} = &i \frac{\epsilon(\omega_d)}{4} \frac{k_d}{t_d} \left(\mathcal{F}^{s\mp c^*}_d t_st_c^*I_sI_c^* + \mathcal{F}^{u c_2^*}_d t_u t_{c_2}^* I_u I_{c_2}^* \right) \\
+ &i \frac{\xi(\omega_d)}{8}k_dI_d\left(\mathcal{F}^{ddd^*}_d\lvert t_d\rvert^2 \lvert I_d\rvert^2 + 2\mathcal{F}^{d\pm a \pm a^*}_d\lvert t_a\rvert^2\lvert I_a \rvert^2 + 2\mathcal{F}^{d\mp c \mp c^*}_d\lvert t_c\rvert^2\lvert I_c \rvert^2 + \sum_{m\in\{s,i,u,c_2\}}2\mathcal{F}^{dmm^*}_d\lvert t_m\rvert^2\lvert I_m \rvert^2\right) \\
+ &i \frac{\xi(\omega_d)}{8}\frac{k_d}{t_d}\left( \mathcal{F}^{u\mp c^*\mp c^*}_d t_u t_c^{*2} I_u I_c^{*2} + 2\mathcal{F}^{\mp c s c_2^*}_d t_c t_s t_{c_2}^* I_c I_s I_{c_2}^* + 2\mathcal{F}^{\pm a i^* \mp c^*}_d t_a t_i^* t_c^* I_a I_i^* I_c^*\right)
\label{eq:CMEd}
\end{aligned}\\
&\begin{aligned}
\frac{\partial I_u}{\partial x} = &i \frac{\epsilon(\omega_u)}{4} \frac{k_u}{t_u} \left( \mathcal{F}^{s\mp c}_u t_st_c I_sI_c + \mathcal{F}^{c_2 d}_u t_{c_2} t_d I_{c_2} I_d \right)\\
+ &i \frac{\xi(\omega_u)}{8}k_uI_u\left(\mathcal{F}^{uuu^*}_u\lvert t_u\rvert^2 \lvert I_u\rvert^2 + 2\mathcal{F}^{u\pm a \pm a^*}_u\lvert t_a\rvert^2\lvert I_a \rvert^2 + 2\mathcal{F}^{u\mp c \mp c^*}_u\lvert t_c\rvert^2\lvert I_c \rvert^2 + \sum_{m\in\{s,i,d,c_2\}}2\mathcal{F}^{umm^*}_u\lvert t_m\rvert^2\lvert I_m \rvert^2\right) \\
+ &i \frac{\xi(\omega_u)}{8}\frac{k_u}{t_u}\left( \mathcal{F}^{d \mp c\mp c}_u t_d t_c^2 I_d I_c^2 + 2\mathcal{F}^{c_2 \mp c^* s}_u t_{c_2} t_c^* t_s I_{c_2} I_c^* I_s + 2\mathcal{F}^{\pm a i^* \mp c}_u t_a t_i^* t_c I_a I_i^* I_c\right)
\label{eq:CMEu}
\end{aligned}\\
&\begin{aligned}
\frac{\partial I_{c_2}}{\partial x} = &i \frac{\epsilon(\omega_{c_2})}{4} \frac{k_{c_2}}{t_{c_2}} \left( \frac{1}{2}\mathcal{F}^{\mp c\mp c}_{c_2} t_c^2 I_c^2 + \mathcal{F}^{u d^*}_{c_2} t_u t_d^* I_u I_d^*\right)\\
+ &i \frac{\xi(\omega_{c_2})}{8}k_{c_2}I_{c_2}\left(\mathcal{F}^{{c_2}{c_2}{c_2}^*}_{c_2}\lvert t_{c_2}\rvert^2 \lvert I_{c_2}\rvert^2 + 2\mathcal{F}^{{c_2}\pm a \pm a^*}_{c_2}\lvert t_a\rvert^2\lvert I_a \rvert^2 + 2\mathcal{F}^{{c_2}\mp c \mp c^*}_{c_2}\lvert t_c\rvert^2\lvert I_c \rvert^2 + \sum_{m\in\{s,i,d,u\}}2\mathcal{F}^{{c_2}mm^*}_{c_2}\lvert t_m\rvert^2\lvert I_m \rvert^2\right)\\
+ &i \frac{\xi(\omega_{c_2})}{8}\frac{k_{c_2}}{t_{c_2}} \left( 2\mathcal{F}^{\mp c s d^*}_{c_2} t_c t_s t_d^* I_c I_s I_d^* + 2\mathcal{F}^{u s^* \mp c}_{c_2} t_u t_s^* t_c I_u I_s^* I_c \right)
\label{eq:CMEc2}
\end{aligned}
\end{align}
where we explicitly added $\pm$ or $\mp$ in front of the terms whose sign changes between the forward ($+$ or $-$) and backward ($-$ or $+$) response. Here,
\begin{equation}
    \mathcal{F}^{\pm m\pm n}_{\pm p} = \frac{(e^{\pm ik_m x} + \Tilde{\Gamma}_m e^{\mp ik_mx})(e^{\pm ik_n x} + \Tilde{\Gamma}_n e^{\mp ik_nx})}{e^{\pm ik_p x} - \Tilde{\Gamma}_p e^{\mp ik_px}}
\end{equation}
and
\begin{equation}
    \mathcal{F}^{\pm m\pm n \pm p}_{ \pm q} = \frac{\left(e^{\pm ik_m x} + \Tilde{\Gamma}_m e^{\mp ik_mx}\right)\left(e^{\pm ik_n x} + \Tilde{\Gamma}_n e^{\mp ik_nx}\right)\left(e^{\pm ik_p x} + \Tilde{\Gamma}_p e^{\mp ik_px}\right) }{e^{\pm ik_q x} - \Tilde{\Gamma}_q e^{\mp ik_qx}}
\end{equation}
are the terms that account for the linear phase mismatch and for reflections for the 3WM and 4WM processes, respectively. In other words, when there is no reflection and when the linear process is phase matched, the corresponding term is equal to unity. For example, if $k_a=k_s+k_i$, and $\Tilde{\Gamma}_a=\Tilde{\Gamma}_s=\Tilde{\Gamma}_i=0$, we have $\mathcal{F}^{si}_{\pm a} = 1$. Note that these CME include soft pump effects.

\subsection{The phase-matching condition}
\label{sec:pmcond}

The 4WM mixing terms (generated by $\xi\neq0$), affect the ideal, trivial 3WM phase-matching (PM) conditions we would otherwise have for the parametric amplification (PA) and frequency conversion (FC) processes, $k_a=k_s+k_i$ and $k_d=k_s-k_c$ (or $k_u=k_s+k_c$ for the up conversion), respectively. To find the modified PM conditions, one has to look for solutions of the CME in the strong pump case, where $\lvert I_a\rvert \gg \{\lvert I_s\rvert, \lvert I_i\rvert\}$ on the one hand, and $\lvert I_c\rvert \gg \{\lvert I_s\rvert, \lvert I_d\rvert\}$ (or  $\lvert I_c\rvert \gg \{\lvert I_s\rvert, \lvert I_u\rvert\}$ for the up conversion) on the other hand. The two PM conditions can be solved for independently. Including the effect of reflections, one can show that \cite{Kern2023reflection_s}:
\begin{equation}
    \Delta\beta_{PA} = \Delta k_{PA} + \chi_{PA}(1+\Gamma_a^2)(k_a-2k_s-2k_i) = k_a-k_s-k_i + \chi_a(1+\Gamma_a^2)(k_a-2k_s-2k_i) = 0,
    \label{eq:PMPA}
\end{equation}
is the 3WM PM condition for the PA process, where 
\begin{equation}
    \chi_a = \frac{\xi \lvert I_a(x=0) \rvert^2}{8} = \frac{\xi \lvert I_{a0} \rvert^2}{8}
\end{equation}
is the PA pump self-phase modulation. Similarly,
\begin{equation}
    \Delta\beta_{FC} = \Delta k_{FC} + \chi_c(1+\Gamma_c^2)(k_c+2k_d-2k_s) = k_c+k_d-k_s + \chi_{FC}(1+\Gamma_c^2)(k_c+2k_d-2k_s) = 0,
\label{eq:PMFC}
\end{equation}
is the 3WM PM condition for the FC process, where
\begin{equation}     
    \chi_{c} = \frac{\xi \lvert I_c(x=0) \rvert^2}{8} = \frac{\xi \lvert I_{c0} \rvert^2}{8}
\end{equation}
is the FC pump self-phase modulation. In practice, the two PM conditions, Eqs.\,\ref{eq:PMPA} and \ref{eq:PMFC}, are only satisfied for two selected triplets of frequencies $\{\omega_a,\omega_s,\omega_i\}$ and $\{\omega_c,\omega_s,\omega_d\}$, respectively, because the line is designed to be dispersive, see Sec.\,\ref{sec:dispeng}. Varying the frequencies of the pumps allows us to move the signal, idler and down-converted tone frequencies at which PM takes place, but we optimize the TWPAC design to obtain the PM conditions with signal and idler tones detuned by $\Delta\omega_\mathrm{PM}=1$\,GHz. It is a somewhat arbitrary choice (and the design does not sensitively depend on it), which ensures us to have a wideband amplifier response.

\subsection{The amplification and conversion gain}

We numerically solve the CME (Eqs.\,\ref{eq:CMEa}-\ref{eq:CMEc2}) in the forward and backward directions, using a set of initial conditions, where we specify an input PA pump amplitude $I_{a0}$, an input FC pump amplitude $I_{c0}$, and a small signal $I_{s0}\ll\{I_{a0},I_{c0}\}$. Any other mode amplitude is initially set to zero. Once the response has been solved, we obtain the current amplitude for each mode in the forward ($+$) and backward ($-$) directions, as a function of the position within the line. We then compute the gain response for all the modes, and in particular for the signal we have:
\begin{equation}
    G_s^\pm(x) = \left(\frac{I_s^\pm(N)}{I_{s0}}\right)^2 T_s,
\end{equation}
where $N$ is the total number of cells within the TWPAC and where
\begin{equation}
    T_s = (1-\Gamma_s^2)^2 \lvert t_s \rvert^2
\end{equation}
is the (power) transmission coefficient, that accounts for the possible impedance mismatch between the TWPAC and its environment \cite{Kern2023reflection_s}.
   
\section{The TWPAC linear response: dispersion engineering}
\label{sec:dispeng}

Dispersion engineering (DE) is the technique by which we accomplish phase-matching (PM) for the PA and FC processes (Eqs.\,\ref{eq:PMPA} and \ref{eq:PMFC}, respectively). In a linear transmission line, a pure 3WM process is naturally phase-matched, however, as shown in Eqs.\,\ref{eq:PMPA} and \ref{eq:PMFC}, the presence of the higher order nonlinearity (the Kerr) affects the PM. Also, in our case the line is naturally weakly dispersive, due to the capacitances in parallel with each JJ. It creates the so-called the chromatic dispersion, a low-Q resonance at the plasma frequency $\omega_p/2\pi\approx40$\,GHz, which affects the dispersion relation at lower frequencies. That is why we need to engineer additional resonances: by placing the pump frequencies close to these resonances, we retrieve the PM conditions.

We aim to have $f_a=14$\,GHz (the 3WM PA pump frequency), so that the amplification band is centered around $f_a/2=7$\,GHz. We therefore engineer a resonant phase-matching (rpm) circuit: a resonator to ground (see Fig.\,\ref{fig:concept}b), composed of an inductance $L_\mathrm{rpm}$ in parallel with a capacitor $C_\mathrm{rpm}$, with resonant frequency $f_\mathrm{rpm}=13.51$\,GHz, slightly lower than the desired $f_a$. It is inserted every 6 cells, and its mode impedance is $20$\,\ohm{}, both of which are a compromise between the resonance quality factor (dictating how sensitive the pump placement will be) and the resonator footprint on chip, which depends on the amorphous silicon thickness ($t=300$\,nm) and dielectric constant ($\epsilon_r = 9.2$).

\begin{figure*}[h]
	\includegraphics[scale=0.95]{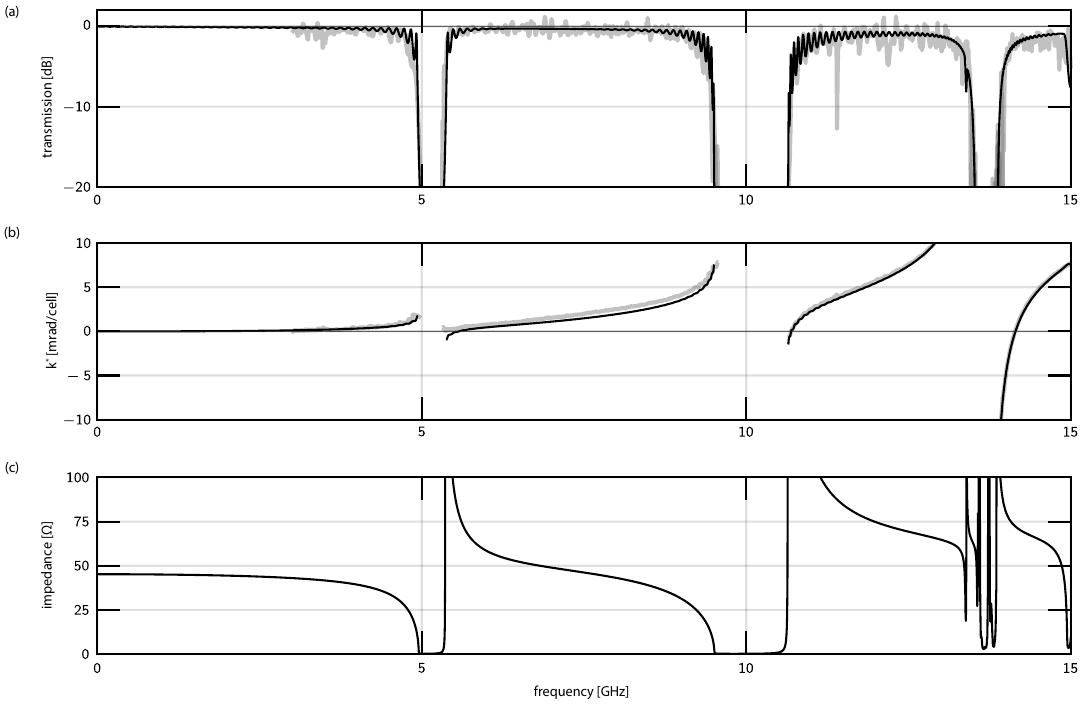}
	\caption{The TWPAC unbiased linear response, using the design parameters presented in Tab.\,\ref{tab:DEparams}. (a) The transmission as a function of frequency (black), calculated from the ABCD matrices of the TWPAC's components, quantitatively agrees to the measured transmission (gray). (b) Similarly, the calculated nonlinear part of the wavenumber $k^*$ (black) quantitatively agrees with the measured one (gray). (c) The calculated TWPAC's impedance as a function of frequency has been optimized to be close to $50$\,\ohm between $6$ and $8$\,GHz.}
\label{fig:TWPAClin}
\end{figure*}

We also aim to have $f_c=4.7$\,GHz (the FC pump frequency), so that the up- and down- converted tones are far-enough detuned from the bandwidth of interest. Concurrently, we aim to suppress the second harmonic at $2f_c$ by placing it into a stopband. Using resonators to create resonances at such low frequencies is inconvenient, because it would lead to having very big capacitor pads, which would complicate the TWPAC layout. That is why we opted for periodically loading the line: we sinusoidally modulate the value of the ground coupling capacitor $C_c(x)$ along the line, \ie{} we modulate the line's impedance. We overlap two modulations, one that creates the stopband and $f_c$, and one creates the stopband around $2f_c$:
\begin{equation}
    Z_\mathrm{pl} = Z_m\left(1 + \delta_c \cos{\frac{2\pi}{N_0}x} + \delta_{c2} \cos{\frac{4\pi}{N_0}x}\right).
\end{equation}
Here, $Z_m$ is a mean impedance, which adjusted to $47$\,\ohm{} in order to have the TWPAC impedance be as close to $50$\,\ohm{} as possible in the bandwidth of interest, around $7$\,GHz. It is an effect of the first periodic loading at $4.7$\,GHz, which slightly increases the nearby impedances above resonance. $N_0$ is the number of cells within a the smallest repetition pattern within the line, composed of both modulations and of the rpm circuits, which we call a supercell. In other words, we chose $N_0$ so that the periodicity of all the repetition patterns are commensurate within one supercell. We adjusted both modulation depths, $\delta_c$ and $\delta_{c2}$, to achieve PM for the FC process, and at the same time have $2f_c$ fall into the second stopband.

\begin{figure}[h]
	\includegraphics[scale=0.95]{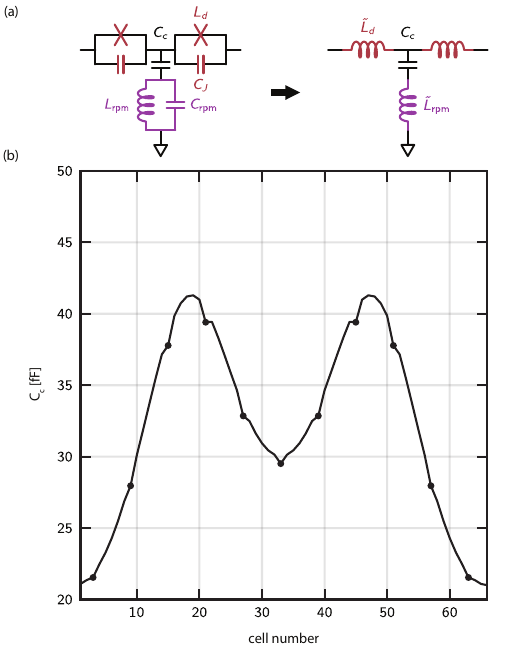}
	\caption{Design of the ground coupling capacitor $C_c$. (a) the rpm circuit (purple) as well as the Josephson capacitor $C_J$ (red) slightly affect the design of $C_c$ (black). Both the rpm circuit and the static part of the capacitively shunted Josephson junction can be modeled (below their resonance) as effective inductors, $\Tilde{L}_\mathrm{rpm}$ and $\Tilde{L}_d$ respectively. (b) The value of $C_c$ within a supercell containing $66$ cells is modulated according to Eq.\,\ref{eq:Cc}. The black dots indicate the presence of an rpm circuit.}
\label{fig:Cc}
\end{figure}

The junction's parallel capacitance $C_J$ as well as the rpm circuit slightly affect the value of $C_c$, see Fig.\,\ref{fig:Cc}a. In fact, the impedance of a symmetric unit cell can be calculated from its ABCD matrix coefficients: $Z_\mathrm{uc}=\sqrt{B_\mathrm{uc}/C_\mathrm{uc}}$, which in the case of the TWPAC gives:
\begin{equation}
    Z_\mathrm{uc} = \sqrt{\frac{\Tilde{L}_d}{C_c}\left[2-C_c\omega^2(2\Tilde{L}_\mathrm{rpm}-\Tilde{L}_d)\right]}, 
    \label{eq:Z}
\end{equation}
where $\Tilde{L}_d$ is given by Eq.\,\ref{eq:tildeLd}, and where
\begin{equation}
    \Tilde{L}_\mathrm{rpm} = \frac{L_\mathrm{rpm}}{1-L_\mathrm{rpm}C_\mathrm{rpm}\omega^2}
\end{equation}
if the cell contains a resonator, or $\Tilde{L}_\mathrm{rpm} = 0$ if it does not. Therefore, inverting Eq.\,\ref{eq:Z} and taking $Z_\mathrm{uc}=Z_\mathrm{pl}(x)$ we obtain:
\begin{equation}
    C_c(x) = \frac{2\Tilde{L}_d}{Z_\mathrm{pl}(x)^2 + \Tilde{L}_d\omega^2(2\Tilde{L}_\mathrm{rpm} + \Tilde{L}_d)}.
    \label{eq:Cc}
\end{equation}

Figure \ref{fig:Cc}b shows $C_c(x)$, Tab.\,\ref{tab:DEparams} summarizes all the TWPAC design parameters and Fig.\,\ref{fig:TWPAClin} shows the TWPAC linear response, calculated by cascading the ABCD matrices of its various components. The overall TWPAC transmission is calculated from the overall ABCD matrix coefficients:
\begin{equation}
    S_\mathrm{21} = \frac{2}{A + B/Z_0 + CZ_0 + D},
\end{equation}
where $Z_0=50$\,\ohm{} is the TWPAC's environment impedance. The overall TWPAC nonlinear wave number is $k^*=k-\omega/v_p$, where $v_p=1/\sqrt{L_d\braket{C_c(x)}_x}$ is the TWPAC phase velocity (at $\omega=0$) with $\braket{}_x$ the space averaging, and where
\begin{equation}
    k = \frac{1}{N}\mathrm{unwrap}\left[\Im\left(\arccosh{\frac{A+D}{2}}\right)\right].
\end{equation}
The overall TWPAC impedance is $Z=\sqrt{B/C}$.

\begin{table}[] 
        \centering
        \begin{tabular}{lll}
        \hline
        \hline        
        parameter & value \\
        \hline        
        Junctions critical current & $I_c\approx5$\,\microamp\\                        
        Junctions plasma frequency & $\omega_p/2\pi \approx 40$\,GHz\\
        Josephson capacitor & $C_J=240.5$\,fF \\        
        mean periodic loading impedance & $Z_m=47$\,\ohm\\
        fundamental impedance modulation depth & $\delta_c=0.1$\\
        second harmonic impedance modulation depth & $\delta_{c2}=0.12$\\
        rpm spacing & $\Delta_\mathrm{rpm}=6$\,cells\\
        rpm circuit frequency & $f_\mathrm{rpm} = 13.51$\,GHz\\
        rpm circuit mode impedance & $Z_\mathrm{rpm} = 20$\,\ohm\\
        rpm circuit coupled quality factor & $Q_\mathrm{crpm} = 578$\\
        rpm circuit inductance & $L_\mathrm{rpm} = 230$\,pH\\
        rpm circuit capacitance & $C_\mathrm{rpm} = 557$\,fF\\
        dc biasing current & $I_d = 1.5$\,\microamp\\
        static inductance under dc bias (at $\omega_a/2$) & $\Tilde{L}_d(\omega_a/2)=71.4$\,pH\\
        first order nonlinearity (at $\omega_a$) & $\epsilon(\omega_a)=0.075$\,\microamp$^{-1}$\\
        second order nonlinearity (at $\omega_a$) & $\xi(\omega_a)=0.033$\,\microamp$^{-2}$\\
        total number of cells & $N=2640$ \\
        number or supercells & $N_\mathrm{sc}=40$\\
        number of cells per supercell & $N_0=66$\\
        mean coupling capacitance to ground & $\braket{C_c(x)}_x=32.2$\,fF\\
        phase velocity (at $\omega=0$) & $v_p=670$\,cell.ns$^{-1}$\\
        cell length & $l_\mathrm{uc}=7$\,\micron\\        
        design signal-to-idler detuning at PM & $\Delta\omega_\mathrm{PM}=1$\,GHz\\
        PA pump frequency & $\omega_a/2\pi=14.5$\,GHz \\
        PA pump chip-input power & $P_a=-73.4$\,dBm \\
        FC pump frequency & $\omega_c/2\pi=4.7$\,GHz \\
        FC pump chip-input power & $P_a=-72.2$\,dBm \\
        phase-matched signal frequency & $\omega_s/2\pi = 7.65$\,GHz\\
        phase-matched idler frequency & $\omega_i/2\pi = 6.65$\,GHz\\
        phase-matched down-converted frequency & $\omega_d/2\pi = 2.95$\,GHz\\
        signal chip-input power & $P_a=-133$\,dBm \\  
        a-Si thickness & $t=300$\,nm \\
        a-Si dielectric constant & $\epsilon_r=9.2$\\
        loss tangent & $\tan\delta=4\times10^{-4}$ \\
        \hline
        \hline
        \end{tabular} 
        \caption{List of the TWPAC design parameters.}
        \label{tab:DEparams}
\end{table}

\section{Losses in the TWPAC}

Our original TWPAC design did not account for any microwave loss. However, upon comparing the TWPAC experimental linear response (\ie{} unpumped, unbiased) to the theoretical one, see Fig.\,\ref{fig:TWPAClin}, a quantitative agreement was found when assuming a loss tangent $\tan\delta=4\times10^{-4}$, a value that is in agreement with previous results of devices fabricated via the same process \cite{lecocq2017nonreciprocal_s}. Thus, the harmonic balance simulations presented in the main text (Fig.\,\ref{fig:design}) as well as those presented throughout the supplementary information all assume $\tan\delta=4\times10^{-4}$. Modeling this loss channel as an admittance in parallel with the coupling capacitance $C_c$ and with the rpm capacitance $C_\mathrm{rpm}$, we account for this loss by performing the transformation:
\begin{equation}
    C_x \rightarrow C_x(1-i\tan\delta),
\end{equation}
where $C_x\in\{C_c,C_\mathrm{rpm}\}$.

It generates a microwave loss that is current bias dependent. In fact, in general the propagation constant $\gamma$ for a lossy transmission line, whose loss is modeled by a resistance $R$ in series with the inductor $L$ and an admittance $G$ in parallel with the capacitor $C$ is \cite{pozar2011microwave_s}
\begin{equation}
    \gamma = \alpha + i\beta = \sqrt{(R + i\omega L)(G + i\omega C)},
\end{equation}
where $\{\alpha,\beta\}\in\mathcal{R}$. When $\alpha\neq0$, signals attenuate along the line. Assuming that $R=0$, we can eliminate $\beta$ and solve for $\alpha$ to find:
\begin{equation}
    \alpha = \frac{1}{2}G\sqrt{\frac{L}{C}} = \frac{GZ}{2}.
\end{equation}
In our situation with the TWPAC, when increasing the current bias we effectively increase $L$, which increases the attenuation constant $\alpha$ (note that $S_{21}=e^{-2\alpha N}$ and $\arg{S_{21}}=-\beta N$). In other words, when increasing the current bias we increase the TWPAC effective length, and therefore we increase its attenuation accordingly.

\begin{figure*}[h]
	\includegraphics[scale=0.95]{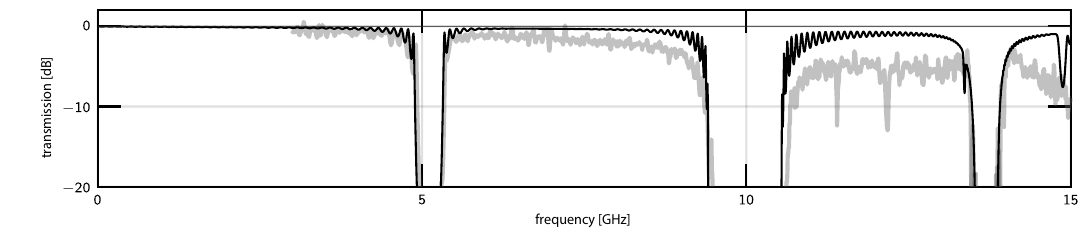}
	\caption{The TWPAC transmission as a function of frequency, when biased with a $1$\,\microamp{} dc current, calculated from the ABCD matrices (black) and measured (gray).}
\label{fig:TWPAClinbiased}
\end{figure*}

However, it does not account for all the loss observed when the TWPAC is current biased, see Fig.\,\ref{fig:TWPAClinbiased}. This additional loss channel could be due to a bias dependent quasi-particle current, that would be modeled as a current-dependent resistor $R_\mathrm{qp}$ in parallel with each JJ, with $R_\mathrm{qp}(I_d=0)=0$. The origin of this additional loss channel remains unclear, so we chose to not account for it in the TWPAC modelling.

The time-domain simulations do not account for any loss, neither dielectric loss nor this additional loss channel, because WRspice does not straightforwardly include the possibility of defining complex capacitors, nor does it allow to define a current-dependent resistor.

\section{Simulation when matching to the up conversion process}
\label{sec:upconv}

\begin{figure}[h]
	\includegraphics[scale=0.95]{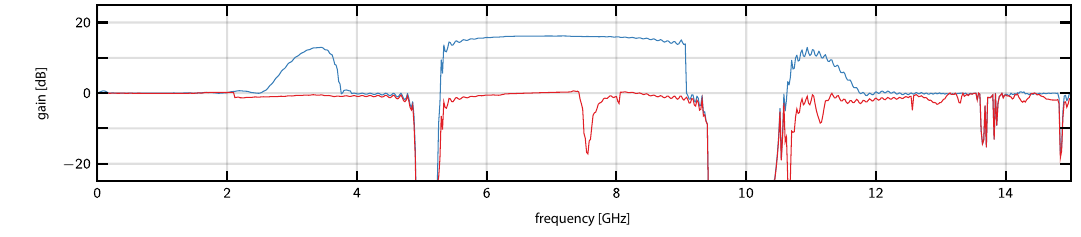}
	\caption{The TWPAC transmission as a function of frequency, when biased with a $1$\,\microamp{} dc current, and pumped with a forward PA pump at frequency $f_a=14.34$\,GHz and power $P_a=-71$\,dBm, and backward FC pump at frequency $f_c=3.15$\,GHz and power $P_c=-76$\,dBm. Here, the CME include all the 3WM and 4WM terms, presented in Eqs.\,\ref{eq:CMEa} to \ref{eq:CMEc2}.}
\label{fig:TWPACmodifiedCME}
\end{figure}

When designing the TWPAC, we chose a FC pump frequency $\omega_c$ that phase-matches the down-conversion process: $\omega_d = \omega_s - \omega_c$, and we did not include the effect of the 4WM processes. For a signal frequency $\omega_s/2\pi = 7.65$\,GHz, it yields  $\omega_c/2\pi=4.7$\,GHz, which dictates the position of the second stopband, placed to prevent the propagation of the second harmonic of this pump tone (at $2\omega_c$). However, in practice the TWPAC only demonstrates backward isolation when pumped at a lower FC pump frequency, around $\omega_c/2\pi=3.15$\,GHz, which corresponds to matching the up-conversion process, $\omega_u = \omega_s + \omega_c$.  In fact, if we solve the coupled-mode equations with $\omega_c/2\pi=3.15$\,GHz, and if we include the 4WM processes (see Eqs.\,\ref{eq:CMEa} to \ref{eq:CMEc2}), we find a backward transmission profile that is qualitatively similar to the one obtained experimentally, as well as the one obtained from the time-domain simulations, see Fig.\,\ref{fig:TWPACmodifiedCME}.

The inability to pump the down-conversion process may be due to the fact that the FC pump frequency that matches this process is too close to the first stopband, and therefore experiences too much of an impedance mismatch when entering the TWPAC (see Fig.\,\ref{fig:TWPAClin}c). This mismatch would (i) prevent the FC pump from efficiently entering the TWPAC and (ii) make the TWPAC resonant at this particular frequency, preventing the directional FC process. Also, it is possible that the FC pump frequency that matches the down-conversion process generates too much of 4WM amplification in the $6$ to $8$ GHz band of interest.

\section{The time-domain simulations}

We closely follow the work of Gaydamachenko et al. \cite{gaydamachenko2022numerical_s} to perform the time-domain simulations. We use WRspice \cite{WRspice_s}, and the circuit netlist is defined from the circuit parameters presented in Tab.\,\ref{tab:DEparams}, and using the periodically varying coupling capacitance to ground $C_c$ presented in Fig.\,\ref{fig:Cc}. The dc current bias is modeled with a dc current source placed between the first and last nodes of the TWPAC and the rf pumps are modeled with sinusoidal sources. The TWPAC environment is set using two $Z_0=50$\,\ohm{} shunt resistors, at its input and output. We perform two sets of simulations: when simulating the forward response, the PA pump is placed at the TWPAC input, together with a small input signal (sinusoidal source with amplitude $0.05$\,\microamp), while the FC pump is placed at the TWPAC output. When simulating the backward response, the PA and FC pumps are swapped. We sweep the signal frequency between $20$\,MHz and $12$\,GHz, stepping it every $20$\,MHz. For each signal frequency we perform a forward and a backward time-domain simulation.

For each simulation, we measure the voltage to ground as a function of time, $V_\mathrm{in}(t)$ and $V_\mathrm{out}(t)$, at the first and last node respectively, as well as the time currents entering ($I_\mathrm{in}(t)$) and exiting ($I_\mathrm{out}(t)$) the TWPAC. After Fourier transforming these four time vectors, we form the transmission scattering parameter:
\begin{equation}
    S_\mathrm{21} = \frac{V_\mathrm{out}(\omega_s) + Z_0I_\mathrm{out}(\omega_s)}{V_\mathrm{in}(\omega_s) + Z_0I_\mathrm{in}(\omega_s)},
\end{equation}
where $\omega_s$ is the signal frequency. We then calculate the gain in dB as $G_\mathrm{dB}=10\log\lvert S_\mathrm{21} \rvert^2$.

Note that we chose the RSJ junction model of WRspice (the so-called \texttt{level=1}), and set the shunt resistance to zero (using the option \texttt{rtype=0}). Otherwise, the WRspice junction model assumes a default shunt resistance \cite{WRspice_s}.




\end{document}